%% file: 0_main.tex
\documentclass[acmsmall]{acmart}
\usepackage[utf8]{inputenc}

\PassOptionsToPackage{table,xcdraw}{xcolor} 

\usepackage{multirow}
\usepackage{subcaption}
\usepackage{xcolor,colortbl}
\definecolor{gray}{rgb}{0.1,0.1,0.1}
\usepackage[T1]{fontenc}
\usepackage{graphicx}
\usepackage{tabularx}
\usepackage{longtable}
\usepackage{enumitem}
\usepackage{booktabs}
\usepackage{setspace}

\AtBeginDocument{%
  \providecommand\BibTeX{{%
    \normalfont B\kern-0.5em{\scshape i\kern-0.25em b}\kern-0.8em\TeX}}}

\setcopyright{cc}
\setcctype{by}
\acmJournal{PACMHCI}
\acmYear{2025} \acmVolume{9} \acmNumber{2} \acmArticle{CSCW151} \acmMonth{4} \acmPrice{15.00} \acmDOI{10.1145/3711049}

\acmSubmissionID{123-A56-BU3}

\input{sections/header}

\begin{document}
\title{Privacy as Social Norm: Systematically Reducing Dysfunctional Privacy Concerns on Social Media}

\author{JaeWon Kim}
\email{jaewonk@uw.edu}
\orcid{0000-0003-4302-3221}
\affiliation{%
  \institution{Information School, University of Washington}
  \country{USA}
}

\author{Soobin Cho}
\authornote{Both authors contributed equally to this research.}
\orcid{0000-0002-4832-208X}
\email{soobin30@uw.edu}
\affiliation{%
  \institution{Human Centered Design \& Engineering, University of Washington}
  \country{USA}
}

\author{Robert Wolfe}
\authornotemark[1]
\email{rwolfe3@uw.edu}
\orcid{0000-0001-7133-695X}
\affiliation{%
  \institution{Information School, University of Washington}
  \country{USA}
}

\author{Jishnu Hari Nair}
\email{jishnu@uw.edu}
\orcid{0009-0000-3490-2465}
\affiliation{%
  \institution{University of Washington}
  \country{USA}
}

\author{Alexis Hiniker}
\email{alexisr@uw.edu}
\orcid{0000-0003-1607-0778}
\affiliation{%
  \institution{Information School, University of Washington}
  \country{USA}
}

\renewcommand{\shortauthors}{JaeWon Kim, Soobin Cho, Robert Wolfe, Jishnu Hari Nair, \& Alexis Hiniker}

\received{January 2024}
\received[revised]{July 2024}
\received[accepted]{October 2024}

\begin{abstract}
Through co-design interviews ($N=19$) and a design evaluation survey ($N=136$) with U.S. teens ages 13-18, we investigated teens' privacy management on social media. Our study revealed that 28.1\% of teens with public accounts and 15.3\% with private accounts experience \textit{dysfunctional fear}, that is, fear that diminishes their quality of life or paralyzes them from taking necessary precautions. These fears fall into three categories: fear of uncontrolled audience reach, fear of online hostility, and fear of personal privacy missteps. While current approaches often emphasize individual vigilance and restrictive measures, our findings show this can paradoxically lead teens to either withdraw from beneficial social interactions or resign themselves to accept privacy violations, viewing them as inevitable. Drawing on teen input, we developed and evaluated ten design prototypes that emphasize empowerment over fear, system-wide explicit emphasis on privacy, clear privacy norms, and flexible controls. Survey results indicate teens perceive these approaches as effectively reducing privacy concerns while preserving social benefits. Our findings suggest that platforms will be more likely to protect teens' privacy and less likely to manufacture unnecessary fear if they include designs that minimize the impact on other users, have low trade-offs with existing features, require minimal user effort, and function independently of community behavior. Such designs include: 1) alerting users about potentially unintentional personal information disclosure and 2) following up on user reports.
\end{abstract}



\maketitle

\input{sections/1_introduction}
\input{sections/2_related-work}
\input{sections/3_method}
\input{sections/4_results}
\input{sections/5_discussion}
\input{sections/6_conclusion}

\begin{acks}
The authors would like to acknowledge the CERES network, which provided support for this work. We additionally thank the anonymous reviewers for their detailed feedback and the participants for sharing their thoughts. We truly appreciate all the help. Alexis Hiniker is a special government employee for the Federal Trade Commission. The content expressed in this manuscript does not reflect the views of the Commission or any of the Commissioners.
\end{acks}

\bibliographystyle{ACM-Reference-Format}
\bibliography{references}

\input{sections/7_appendix}

\end{document}

%% file: sections/header.tex
\input{formatting/published-packages} 

\input{formatting/dev-packages} 
\input{formatting/dev-commands} 

%% file: formatting/published-packages.tex
\usepackage{xspace}     
\usepackage{xpunctuate} 

%% file: formatting/dev-packages.tex
\usepackage{xargs}      
\usepackage{soul}       
\usepackage{color}      

%% file: formatting/dev-commands.tex
\definecolor{LIGHTPINK}{RGB}{237,157,202}
\definecolor{LIGHTRED}{RGB}{210,121,121}
\definecolor{LIGHTORANGE}{RGB}{230,170,50}
\definecolor{LIGHTGOLD}{RGB}{210,194,121}
\definecolor{LIGHTGREEN}{RGB}{121,210,121}
\definecolor{LIGHTAQUA}{RGB}{121,206,210}
\definecolor{LIGHTBLUE}{RGB}{121,124,210}
\definecolor{LIGHTPURPLE}{RGB}{153,102,255}
\definecolor{RED}{RGB}{178,34,34}
\definecolor{GRAY}{RGB}{166,166,166}
\definecolor{WHITE}{RGB}{255,255,255}

\newcommandx{\jane}[2][1=] 
    {\setulcolor{LIGHTGREEN}{\ul{#1}} \textcolor{LIGHTGREEN}
    {[\textbf{Jane:} #2]}}
\newcommandx{\guest}[3][1=]
    {\setulcolor{LIGHTORANGE}{\ul{#1}} \textcolor{LIGHTORANGE} 
    {[\textbf{#2:} #3]}}


%% file: sections/1_introduction.tex
\section{Introduction}
Due to the pervasive influence of social media, the landscape of teenage social interaction has largely migrated to the online realm~\cite{noauthor_2023-xq}. This shift has led to an escalation in privacy concerns, highlighting the delicate balance between online engagement and personal data protection for teens~\cite{Agha2023-dv, WisniewskiPamela2022}. While debates have persisted about teenagers' attitudes towards privacy, recent studies indicate that they actively engage in various strategies to safeguard their personal information~\cite{zhao2022understanding}. Nevertheless, significant concerns around teen privacy management remain underexplored.

A revelation from a 2019 Pew Research Center study~\cite{auxier2019americans} underscores an important concern around privacy management: 82\% of Americans feel they have little to no control over their personal data, including location tracking. This sentiment is further pronounced among teenagers, who, as a vulnerable group, often feel overwhelmed and powerless in the face of privacy challenges. Research has illuminated several factors contributing to this vulnerability: teens frequently encounter a lack of self-efficacy when navigating privacy issues~\cite{de2020contextualizing}, face conflicting privacy norms among peers~\cite{Trepte2020-fw}, and struggle to find viable solutions to their privacy concerns~\cite{blank2014new}. This confluence of factors can engender a state of \textbf{\textit{dysfunctional fear}}~\cite{jackson2010functional} in teens, which is characterized by heightened anxiety over privacy that paradoxically diminishes their well-being and discourages proactive measures.

In this paper, given the critical need for teen privacy protection in social media, while preventing amplified dysfunctional fear, we investigate the following questions:
\begin{enumerate}
    \item \textbf{RQ1.} What, if anything, do teens fear in their social media experiences? How do these fears affect their social media experiences?
    \item \textbf{RQ2.} How might the design of social media mitigate dysfunctional fear for teens? What are the implications of these designs?
\end{enumerate}

To answer these questions, we conducted a mixed-methods study with a total of 137 13-to-18-year-old participants residing in the U.S. We first conducted co-design interviews with nineteen adolescents to understand privacy concerns they have difficulty addressing and how those affect their social media experience. The participants also shared design suggestions to address their privacy concerns. We then took the design suggestions and created high-fidelity prototypes for ten of the larger design idea themes that we identified from the co-design interview data. We also conducted a design evaluation survey with 72 private account owners and 64 public account owners. 

Our results reveal that teens often experience a vague and persistent fear that remains unaffected by their privacy management strategies on current social media platforms. The fear concerned three overarching themes and seven subthemes, including `fear of time collapse,' `fear of a hostile environment,' and `fear of overstepping privacy norms.' Such fears impact their quality of life, leading them to either withdraw from desired social media activities or become hypervigilant after posting. They also struggled to find sufficient features to address the fears, particularly when they have a public account. Moreover, the overall hostility of social media environments often inhibited their sense of safety. Even when protective options are available, they found that their peers' privacy expectations differ from their own, creating pressure to conform rather than prioritize their preferences. We find that features such as clarifying privacy norms during onboarding and providing default but optional features for privacy protection, such as screenshot blocking, can alleviate their fears while reducing the likelihood of privacy issues.

In this paper, we contribute empirical evidence of dysfunctional fear in the context of teen privacy on social media. We also provide an overview of features and affordances that could effectively reduce both privacy concerns and risks. Furthermore, we identify moments where teens' privacy needs collide with their peers' privacy expectations and suggest methods for platforms to foster a culture where privacy is considered the norm. We hope this work will prompt designers and researchers to consider the adverse effects of alarmism or prevention-focused strategies and, instead, systematically empower adolescent users to protect themselves while fully reaping the benefits of social media.

%% file: sections/2_related-work.tex
\section{Related Work}

\subsection{Teen Privacy Concerns on Social Media}
Social media platforms offer a multitude of benefits, particularly for teens, including the enhancement of social capital and the facilitation of meaningful discourse. Central to these advantages, such as relationship formation, is self-disclosure~\cite{tidwell2002computer, ellison2010little, ellison2011negotiating, taddicken2011uses}. Often, disclosing personal information is a prerequisite to fully leveraging technological tools' potential, such as the different types of social capital benefits~\cite{ellison2007benefits, ellison2011negotiating}. The more users share about themselves, the greater the benefits they can derive from these systems. However, this increased disclosure also escalates the risk of what users perceive as privacy violations~\cite{taddicken2011uses, debatin2011ethics}. It is also unsurprising that safety threats on social media platforms can inhibit users' ability to gain these advantages. Both real and perceived threats can diminish the quality and benefits of interactions on these online social spaces~\cite{redmiles2019just, jozani2020privacy, chen2013problematic, dienlin2016extended, krasnova2009privacy}. This trade-off is particularly unfortunate considering the heavy reliance of modern teens on social media for socialization and identity development~\cite{Hamilton2022-xi, Davis2012-bq}.

Teens generally face many privacy concerns on social media, although their focus might be somewhat different from that of adults. Teens often focus their privacy efforts on evading direct supervision, such as from family members, avoiding interactions with strangers, and circumventing in-person drama with peers~\cite{zhao2022understanding}. Moreover, due to their developmental stage, adolescents are highly self-conscious, often leading to the belief that they are constantly being watched or judged (i.e., \textit{``imaginary audience''}~\cite{elkind1979imaginary}). This belief, coupled with the low audience transparency inherent in social media, heightens teens' concerns about audience scrutiny. In addition, teens' perceptions of privacy evolve as they encounter changing societal norms and expectations, leading to concerns about their digital footprint due to the persistent nature of social media content. Another privacy issue is oversharing, which involves posting excessive personal information or posting too frequently, a concern often developed by observing their peers~\cite{brammer2022oversharing, zhao2022understanding}. In addition, privacy concerns frequently arise due to the challenges in managing networked privacy~\cite{marwick2014networked} on social media. Although teens are quite adept at navigating and configuring settings in networked environments~\cite{zhao2022understanding}, networked privacy by its nature places individuals in social contexts where their privacy can be, and often is, violated, whether intentionally or unintentionally~\cite{hargittai2016can}.

\subsection{Designs and Strategies for Teen Privacy on Social Media}
Contrary to early beliefs that teen privacy revolves around the \textit{``privacy paradox,''} suggesting teens are less concerned with privacy than older individuals~\cite{walrave2012connecting}, it is now understood that valuing privacy is the norm among young people~\cite{zhao2022understanding, blank2014new, Weinstein2022-rh}. Teens experience a conflict between the desire to share information with their friends and the concern of unintended audiences viewing their content~\cite{agosto2017don}. Thus, they actively engage in protecting their privacy through various strategies. These include, but are not limited to, posting in a strategically vague manner to obscure the meaning from those outside their network~\cite{zhao2022understanding}, disengaging from unsafe conversations~\cite{Ali2022-cb}, leveraging ephemeral modes of posting~\cite{xu_automatic_2016, Chiu2021-pj}, gradually decreasing the visibility of posts to mitigate risks associated with digital footprints~\cite{Pias2022-yy, Mohamed2020-fj}, adjusting privacy settings, and practicing self-censorship to prevent \textit{``context collapse''}~\cite{dennen2017context} or \textit{``time collapse''}~\cite{brandtzaeg2018time}. They also control their digital boundaries by filtering follow requests, blocking or removing followers~\cite{Agha2023-mu}, limiting audience reach~\cite{zhao2022understanding}, and creating trusted spaces~\cite{Xiao2020-ce}. Additionally, teens manage their online presence by archiving or deleting content~\cite{Agha2023-mu}, adhering to and respecting implicit networked privacy norms~\cite{zhao2022understanding, peter2011adolescents}, and employing other nuanced measures to maintain their digital privacy. Teens typically employ sophisticated privacy management techniques for their privacy protection needs, even when they believe they ``have nothing to hide''~\cite{adorjan2019student}.

The Computer-Supported Cooperative Work (CSCW) and Human-Computer Interaction (HCI) domains have made significant progress in finding ways to support teens’ privacy protection better. The overall direction of the findings is to move away from designing based on paternalistic, protectionistic, and \textit{``concern-centric''} approaches~\cite{BoydDanah2014ICTS, WisniewskiPamela2022, Jia2015-nt, Agha2023-mu} and toward a \textit{``risk-centric''}~\cite{WisniewskiPamela2022} framework. Teens are turning to social media for connection and relaxation due to decreased public freedom and increased adult supervision in their physical environments. This shift to online spaces allows them to maintain social interactions in a setting that offers more privacy and autonomy, highlighting the need for a balance between guidance and independence~\cite{BoydDanah2014ICTS}. Thus, the risk-centric framework suggests that teens can adopt more adequate privacy protection measures as a result of exposing themselves to increased online disclosures, thereby facing higher susceptibility to risky online interactions~\cite{Jia2015-nt, WisniewskiPamela2022, Wisniewski2018-rc}. Additionally, measures that engage teens more directly and empower them, such as discussions and nudging them toward self-regulation, have been emphasized~\cite{kang2022teens, WisniewskiPamela2022, Agha2023-mu}. Specific designs that support teen privacy include nudging~\cite{Acquisti2017-ez, Masaki2020-wy, Agha2023-mu}, multiple close friends list~\cite{zhao2022understanding}, and preventive measures such as restrictions for perpetrator and sensitivity filter~\cite{Agha2023-mu}. Designs to address adult privacy concerns, such as improving the findability of privacy interfaces, may also be relevant for teens~\cite{Cobb2020-to, Im2023-bb}.

\subsection{Toward a Community-level Privacy Support for Teens on Social Media}
Despite advancements in supporting teen privacy, teens continue to face significant challenges in privacy protection~\cite{razi2020privacy}. A key factor is a reliance on individual teens' conscious efforts for privacy management~\cite{walther2011introduction, Wei2023-zq}, which often falls short when teens perceive their privacy concerns as being beyond their control~\cite{kang2015my}. This perception can arise from a \textit{lack of self-efficacy}~\cite{de2020contextualizing}, \textit{social norms conflicting with individual privacy needs}~\cite{Trepte2020-fw}, or \textit{limited options for adequately addressing privacy issues}~\cite{blank2014new}. Specifically, navigating privacy in networked public spaces requires users to manage technical affordances, interpersonal relationships, and social norms~\cite{zhao2022understanding}, and peer norms significantly influence teen privacy practices~\cite{zhao2022understanding, Trepte2020-fw}. Teens rely on trust that their friends will respect the boundaries of co-owned privacy~\cite{zhao2022understanding}, but social media platforms lack explicit rules and norms around this co-ownership. Consequently, when friendships deteriorate, the fragile agreements about what can and cannot be shared often break down, leading to breaches in privacy~\cite{lampinen2011we}. These challenges align with \textit{Contextual Integrity theory}~\cite{nissenbaum2004privacy}, which posits that privacy is maintained when information flows respect context-specific norms. For teens, the difficulty in managing these norms across various social media contexts can contribute to privacy concerns and feelings of reduced control.

As Emily Weinstein and Carrie James detail in their book~\cite{Weinstein2022-rh}, modern teenagers face unprecedented challenges in managing their online presence. The widespread practice of peers documenting and archiving digital interactions has created an environment of constant surveillance. Young people must navigate the threat of public shaming on social media, where past mistakes can be exposed and lead to widespread criticism or ostracism. The growing trend of saving digital evidence---whether through screenshots, recordings, or saved messages---has intensified these privacy concerns. This culture of perpetual documentation makes adolescents feel increasingly powerless over their personal information and digital reputation. The knowledge that any misstep could be preserved and shared creates significant anxiety among teens, who struggle to maintain autonomy over their online identities in this atmosphere of persistent observation and potential judgment. As such, the complexity and perceived lack of control often leads to \textit{``network defeatism''}~\cite{de2020contextualizing} among teens, a sense of fatalism regarding privacy decisions on social media~\cite{de2020contextualizing, hargittai2016can}. This reduced sense of self-efficacy in privacy control hampers teens' ability to address privacy issues and enhance their social media experiences~\cite{lee2008keeping, jozani2020privacy}. Furthermore, teens are more likely to overlook their vulnerability when they feel less efficacious, leading to disengagement from privacy protection~\cite{chou2023teens}. Adults' well-meaning warnings about the permanence of online actions often add to this disempowerment, instilling a sense of fear and cynicism~\cite{Weinstein2022-rh}.

Echoing themes from fear of crime research, these phenomena underscore the need to understand worries and fears around privacy-related issues on social media. For example, fear of crime research differentiates between \textit{dysfunctional worry}, which degrades the quality of life, and \textit{functional worry}, which fosters precaution and awareness. Fear of crime research has found that only a quarter of worried individuals used their fear constructively, without impacting their life quality~\cite{jackson2010functional}. Fear of cybercrime studies reveal a similar pattern: while fear can be a motivator for protective behaviors, it often restricts activities and thus the quality of life, particularly when the fear response is disproportionate to the actual risk~\cite{brands2022measurement}\footnote{It is important to note, however, that our research does not aim to quantify the actual risks of privacy issues for teens on social media or determine what constitutes dysfunctional fear in this context. Rather, we seek to understand the emotional and psychological mechanisms of (dysfunctional) fear of crime and how they might apply to social media privacy concerns.}. The influence of social environment on fear, as highlighted by the fear of crime research, may also be relevant to social media privacy concerns, suggesting that factors like (perceived) community disorder and (perceived) lack of social support can amplify fear~\cite{lee2019test, greenberg1986fear, lorenc2013fear}. Therefore, in this paper, we argue that addressing social media privacy concerns requires a balance between vigilance and system-wide reassurance~\cite{redmiles2019just}, as well as measures to increase self-efficacy. Following the \textit{``Privacy by Design''} principle~\cite{cavoukian2009privacy}, proactive approaches that integrate privacy into the design are essential. This is particularly important for teens, who are highly susceptible to peer pressure and social norms. However, it is crucial that community-wide privacy features do not merely create a \textit{``safety theater,''}~\cite{redmiles2019just} which focuses solely on reducing fear without real safety, as a false sense of security can be dangerous~\cite{cho2010optimistic, walther2011introduction, acquisti2006imagined, gross2005information, ostendorf2020neglecting}. The aim should be to empower users to move away from dysfunctional fear by transforming their concerns into constructive actions~\cite{jackson2010functional, Chordia-2024-TuningWorldCrime-z}.

%% file: sections/3_method.tex
\section{Method}

\subsection{Co-design Interviews}

\noindent\textbf{Materials and Procedures.} We conducted co-design interviews in two stages: 30-minute entry and 60-minute exit interviews. We explored participants' privacy concerns on social media platforms during entry interviews. We asked generally where they felt most comfortable sharing, with whom they shared, and what (if any) issues or features influenced their sharing habits.

During exit interviews, we asked participants about the barriers (if any) that prevent them from comfortably sharing on social media. To do so, we used the Miro platform~\cite{miro} to create a virtual whiteboard with nine sticky notes, each representing the improvements they seek on social media (and its related subcategories of concerns) identified in our survey. Examples of the larger themes on changes sought were ``prevent posts from being shared out of context,'' while sub-themes of privacy concerns included ``Content being shared with strangers'' and ``Screenshots taken / content saved.'' (The full list of the sticky notes is available in the exit interview template in \href{https://osf.io/8t7cy/}{\url{https://osf.io/8t7cy/}}) We used these to prompt a conversation where participants highlighted the barriers most relevant to them.

Afterward, we conducted a co-design sketching exercise with the participants to dive deeper into these concerns' significance and relevance and to brainstorm potential design solutions. To prompt ideas for designs, we asked questions such as ``What are your perceived social norms for privacy? Can those be flexible or broken? Are these different across different platforms?'' Then, for each design idea the participants came up with, we asked them to sketch and explain their concept. We inquired about the possible benefits and downsides of incorporating the feature and whether such effects would differ across platforms or people.

The first author conducted the semi-structured entry and exit interviews via Zoom, lasting approximately 30 to 40 minutes and 60 to 90 minutes, respectively. Participants were compensated with \$10 and \$20 Amazon gift cards for their participation in the entry and exit interviews.

\vspace{2mm}
\noindent\textbf{Participants and Recruitment.} We reached out to individuals from our previously established participant pool, which was initially formed through purposive and convenience sampling. This sampling involved advertisements on social media platforms such as Instagram and Facebook, specifically targeting individuals aged 13 to 18 in the United States. Additionally, we invited individuals previously involved in or interested in our studies and provided consent to be contacted for future research opportunities. In total, we invited 80 participants to complete a general screening survey for our co-design study. Of the 47 respondents who completed the screener, 22 were chosen to participate in the co-design study. Participants were selected to reflect a broad range of demographic characteristics. Of these 22 invitees, 20 initially participated in the study. However, one participant withdrew after the entry interview. Consequently, 19 participants (demographic details in Appendix \ref{appendix-A}) completed all stages of the co-design study.

\vspace{2mm}
\noindent\textbf{Reflexive Thematic Analysis.} We undertook a reflexive thematic analysis~\cite{braun2019reflecting} of the 38 entry and exit interview transcripts. Our approach was inductive, yet we concentrated on privacy-related aspects---especially privacy fears that participants felt undermined their social media experience---since the interviews encompassed a broader range of topics than could be addressed in a single paper. The analysis was structured into three phases over 2.5 weeks.

In the initial phase, the first author, two co-authors, and the last author engaged in line-by-line coding of two identical transcripts. The initial codes were descriptive, closely reflecting the data. Following this phase, the first author compiled these notes to draft the preliminary codebook. Subsequently, in the second phase, the same team independently coded two additional transcripts using Atlas.ti software~\cite{atlas} using this preliminary codebook and adding new codes as necessary. The first author then reviewed these codes and newly found themes to refine the codebook further.

In the final phase, the primary author coded all transcripts to refine the codebook. The two co-authors then coded 20 of the 38 transcripts, validating the clarity and validity of the codebook and the assigned codes. Throughout each phase, the team convened to discuss higher-level themes, disagreements, or ambiguities related to the codebook. Higher-level themes included ``Fear that teens experience over privacy,'' ``Existing strategies for mitigating fear,'' ``Designs that affect teens' fear,'' and ``Implications of the designs.'' Sub-codes included ``Ripple audience,'' ``Perspective taking,'' ``Prompts for self-regulation,'' and ``Privacy-utility trade-off.'' Each iteration of the codebook, complete with definitions and example quotes for each code, is available in \href{https://osf.io/8t7cy/}{\url{https://osf.io/8t7cy/}}).

\subsection{Design Evaluation Survey}

\noindent\textbf{Materials and Procedures.} For each of the ten design idea themes developed during the inductive coding process of the co-design interviews, the first and fourth authors---both advanced-degree students in technology design programs---created high-fidelity prototypes of ten sample features (See Figure \ref{tab:prototypes} for sample prototypes and \href{https://osf.io/8t7cy/}{\url{https://osf.io/8t7cy/}}) for the entire set of prototypes) that were provided to the survey respondents. These authors selected specific implementations for each design based on their potential to probe the defining characteristics of each design idea.

We incorporated the ten prototypes into a survey, inviting participants to evaluate each prototype in terms of general reactions (e.g., ``What (if anything) do you dislike about this feature?''); usefulness in addressing privacy concerns (e.g., ``With this feature, I would feel less worried about privacy-related problems on my [public/private] account''); likelihood of backfiring (e.g., ``If I use this feature on my public account, I'm concerned it might lead to awkward or uncomfortable situations with my friends or people I know in real life.''); and to answer general questions on demographics; social media use (e.g., ``On which social media platforms do you have (or have you had) a public account? (choose all that apply)''); and privacy concerns related to social media (e.g., ``Thinking about your [public/private] account on [platform], please rate your concern that someone might expose your private conversations with others, either on purpose or by accident''). 

Through these questions, we aimed to understand the extent of teens' dysfunctional fear of privacy risks on social media and assess each feature's potential strengths and weaknesses. The survey's estimated completion time was 38 minutes, with participants receiving \$12 Amazon gift cards for their time. The median completion time was 48.2 minutes, with an interquartile range of 26.9-106.3 minutes.

We hypothesized that the effectiveness of, and thus the reactions to, the features would significantly differ between public and private account owners. We also anticipated they would have different types and extents of privacy concerns. Consequently, we created two versions of the same survey, one for public and another for private account owners. Each version asked participants to consider the features and privacy concerns in the context of the public or private account they most frequently use. We permitted participants to respond to both surveys if they owned both types of accounts. The complete wording of the two versions of surveys can be found in \href{https://osf.io/8t7cy/}{\url{https://osf.io/8t7cy/}}).

\vspace{2mm}
\noindent\textbf{Participants and Recruitment.} We invited 443 selected individuals who met one of two criteria: 1) those who have participated in our past studies and demonstrated sincerity in their participation, and 2) new recruits via an Instagram advertisement who answered the free-response question in the screening survey in good faith. At no point in the recruitment process did we specify a need for teens with specific privacy concerns. Instead, we targeted teens that would ``tell us about [their] social media experience.'' We received a total of 201 responses: 96 from private account owners and 105 from public account owners.

Each response was reviewed, and we excluded those that appeared to be 1) submitted multiple times by an individual, 2) generated by generative AI, or 3) not genuine, such as responses not addressing the question or only offering vague positive comments such as \textit{``I would love this''} without any evidence of having reviewed the prototypes throughout the entire survey. The detailed process and criteria we employed for this process are available in \href{https://osf.io/8t7cy/}{\url{https://osf.io/8t7cy/}}). After this filtering process, data from 72 private account owners and 64 public account owners remained. Among these, 18 participants responded to both surveys. The demographics of the 118 participants (who provided 136 responses) are detailed in Appendix \ref{appendix-A}.

\vspace{2mm}
\noindent\textbf{Evaluating Design Features (One-Sample t-Tests).} We conducted one-sample t-tests on their reactions to each prototype and an overview of all prototypes. These tests assessed whether participants' responses to each survey question significantly deviated from a hypothesized mean of 3, representing ``neither agree nor disagree,'' the midpoint of our survey scale.

\vspace{2mm}
\noindent\textbf{Identifying Dysfunctional Privacy Concerns.} Following the framework of Jackson and Gray~\cite{jackson2010functional}, we classified privacy concerns on social media into three categories: unworried, functional worry, and dysfunctional worry. Our goal was not to determine what constitutes an ``appropriate'' level of worry. Rather, we sought to identify instances of intense worry that diminish the quality of life and are not or cannot be accompanied by effective, constructive actions. Details about our adoption of this categorization algorithm can be found in Appendix \ref{appendix-B}.

\vspace{2mm}
\noindent\textbf{Clustering Open-Ended Feedback.} We analyzed participant feedback about design features using the MPNet-Base-V2 language model~\cite{mpnet-base-v2} from the Sentence Transformers library~\cite{sentence-transformers}, converting responses into 320-dimensional vectors. This process resulted in 136 embeddings for both `likes' and `dislikes.' We then applied k-means clustering~\cite{hartigan1979algorithm} to these embeddings, selecting an optimal number of clusters (between 3 and 7) based on the silhouette coefficient~\cite{dinh2019estimating} for meaningful categorization. Each cluster was labeled by its main theme, with representative quotes provided in Appendix \ref{appendix-C}. These clusters helped us understand the design feedback in a time-efficient and targeted manner; While the clusters are not definitive, we used the clustering results as a guide for stratified and purposive sampling when evaluating the open-ended responses.

\subsection{Ethical Considerations}
Both procedures were approved by our Institutional Review Board (IRB) prior to data collection. We sent consent forms (including parental consent) to participants before the interviews. During the interviews, we summarized the consent form and procedures, allowing participants to ask questions. We obtained both written and verbal consent to inquire about personal social media experiences and to record responses.

Given the sensitivity of the topic, we took additional precautions. We explicitly informed participants that they could decline to answer any questions that made them uncomfortable. Participants were given the option to turn off their cameras. We emphasized prior to starting the interviews that our goal was to understand teens' perspectives, not to judge their experiences or behaviors.

\subsection{Use of Generative AI} We utilized ChatGPT-4~\cite{chatgpt} to generate initial ideas for section headings, prototype names, and theme titles. Typical prompts were: ``[summary of content in the section and long description of the key points we aimed to deliver through the section, written by the authors] What might be some headings for a section that suggests this? Give me a long list.'' We then refined these initial ideas to finalize versions that most accurately delivered what we were looking for. Additionally, the AI helped refine grammar in parts of our paper, draft accessibility texts for figures, and improve the conciseness of our writing. Typical prompts were ``Just fix grammar without summarizing or anything'' and ``Is any part grammatically wrong or awkward?''

%% file: sections/4_results.tex
\section{Results}

\subsection{How Does Fear Affect Teens' Social Media Experiences?}
As shown in prior literature, teens actively took measures to protect their privacy~\cite{zhao2022understanding}. These measures include withholding information, considering their audience's perspective, reviewing and reversing content, assessing boundaries, and developing mental models. Often, they develop these models by observing their peers to understand what constitutes appropriate privacy practices. 

However, as evidenced in prior literature, teenagers also perceive privacy concerns or fears as being beyond their control due to various reasons such as a lack of self-efficacy~\cite{de2020contextualizing}, social norms that conflict with individual privacy needs~\cite{Trepte2020-fw}, or limited options for adequately addressing privacy issues~\cite{blank2014new}. Our survey results also confirmed the presence of dysfunctional fear in teens; the survey revealed that among our survey respondents with a private account, 15.3\% experience dysfunctional fear. Among public account owners, the figure is 28.1\%, almost twice as much.

From our co-design interviews, we identified ten sources of such fears, grouped under three broader themes: `fear of content being shared out of context,' `fear of digital vulnerability,' and `fear of online missteps.' These fears prevent teens from fully utilizing the benefits of social media. We also outlined design approaches that teens suggested to mitigate each type of fear. Although we discussed specific design ideas with specific types of fear, this does not imply that these ideas are exclusively effective for the particular type of fear they are associated with. We associated them based on where they were most frequently mentioned.

\subsubsection{Fear of Uncontrolled Audience Reach} 
The teen participants frequently expressed concerns about their digital content being shared or interpreted outside its original context, driven by the persistent, replicable, and scalable nature of digital media~\cite{Boyd2008-zi} and uncertainties surrounding networked boundaries. To address these fears, participants suggested enhancing visibility and boundary control options, increasing interaction transparency, implementing screenshot controls, and providing flexible content management features such as ephemerality and easy post-deletion.

\vspace{2mm}
\noindent\textbf{Imaginary Audience.} 
Eighteen participants (P02-P19) shared anxieties about potentially having large, unknown audiences on social media. This aligns with previous research findings, highlighting the opaque social media visibility mechanisms~\cite{DeVito2017-by}. These teen participants experienced low audience transparency and felt they had little control over whether their social media content might unintentionally reach unwelcome or unintended audiences. This fear was especially prevalent among those with public accounts. On platforms like YouTube, algorithms can cause content to \textit{``go viral,''} exposing it to \textit{``anyone who uses the platform''} (P04). Even those with private accounts were not immune to this fear, feeling inhibited from sharing due to their \textit{``large following''} (P17). Additionally, even when utilizing features like Instagram's Close Friends Story, some teens felt these lists were too \textit{``general''} (P17) and thus not sufficiently exclusive for certain types of content. In some instances, teens experienced anxiety over not knowing how the audience was engaging with their content, fearing excessive scrutiny or negative judgment from unseen viewers, a phenomenon similar to `stage blindness'---the anxiety of being on stage and unable to see the audience due to bright lights. As P19 succinctly put it, \textit{``We are afraid of the unknown; when you post, you don't really know who has seen it and who hasn't, or who disliked it.''}

One design idea proposed by all nineteen teen participants to alleviate the fear of an imaginary audience involved enhancing visibility and \textit{boundary control} options for public accounts while strengthening these controls for private accounts. They were already actively using various visibility and boundary control mechanisms and strategies on mainstream social media platforms, including rejecting friend requests, removing followers or friends, selective sharing, intimate reconfiguration, and privatizing accounts, as found in prior research. However, such boundary and visibility controls limited the benefits teens sought from social media. For instance, private accounts made it \textit{``harder to meet new people''} (P14), conflicting with their desire to \textit{``constantly networking and expanding [their] audience''}, especially for those pursuing activities like music. As P15 explained, with a public account, you \textit{``you sort of give up the boundary \ldots to other people,''} and must \textit{``accept''} the \textit{``opportunity costs.''} They, therefore, suggested that there is a need for \textit{``a way for people that want to be public to still be able to have some sort of ... knowing what's going on with their content.''} For private accounts, participants desired more nuanced selective sharing options, with various \textit{``groups,''} rather than the limited options, such as the \textit{``only two groups''} that Instagram currently offers. Ideally, these options would cater to different interests, as suggested by P19.

Fifteen participants (P01, P02, P04-P09, P11-P15, P17, P18) suggested increasing \textit{interaction transparency} as another design direction. They particularly desired clarity about \textit{``how people interact with your posts,''} such as identifying viewers and their interactions (P07). For instance, P09 proposed a feature like \textit{``profile views''} that would allow users to see who has viewed their regular posts, similar to Instagram Stories, as \textit{``knowing who sees you''} contributes to a sense of security. Understanding \textit{``how people interact with your posts''} (P07) would not only \textit{``set off warning bells,''} but also provide them with the opportunity to block someone if necessary. This knowledge would offer them \textit{``peace of mind,''} as they would \textit{``prefer to know... rather than be anxious about who is sharing [their] content.''} It would also allow them to recognize and address uncomfortable interactions (P07). Although most participants valued having insight into their content's interactions, some believed that \textit{``sometimes ignorance is better,''} particularly when expected responses to their posts were not forthcoming. They preferred to assume the post was unseen rather than acknowledge it was ignored (P14). Nevertheless, they still acknowledged the importance of interaction transparency and recommended making this feature optional, with the ability to \textit{``toggle it on and off''} (P14).

\vspace{2mm}
\noindent\textbf{Boundary Violations.} 
Among our 19 participants, 17 (P01-P09, P11, P12, P14-P19) expressed concerns that their content, intended for a specific audience, might be intentionally or unintentionally exposed to individuals outside a mutually understood privacy boundary. This trust is sometimes broken with malicious intent, as in cases where \textit{``people take screenshots and send them to individuals whom you didn't intend to see them''} (P07). At other times, the implicit expectations of the shared boundary are breached unintentionally. An example is, \textit{``We'll screenshot the messages we sent to each other and then send them to someone else, which we started doing because we thought it was funny... but then it gets out of hand''} (P06). This is concerning because such breaches of the shared boundary can lead to misunderstandings or misrepresentations when content is viewed outside its original context by those unfamiliar with its background or if a trusted party betrays trust by purposely sharing \textit{``one little snippet''} (P15) of the original content.

Twelve (P01, P02, P06, P07, P09, P11-P13, P15-P18) participants proposed the idea of \textit{screenshot control}, either through notifications or blocking mechanisms. Many felt that screenshot notification would provide an \textit{``extra layer of security''} and a means to \textit{``confront [other users]''} (P14) if necessary. However, some noted drawbacks, such as feeling uncomfortable with Snapchat's feature of notifying others about accidental screenshots (P13). P09 mentioned that knowing someone took a screenshot of their post could be unsettling, leading to thoughts that \textit{``mess up [their] day.''} Some participants preferred that screenshots be entirely blocked, similar to how services like Disney Plus or Netflix prevent screenshots by blacking out the screen (P07). Others suggested that screenshots should \textit{``disappear''} when shared, preventing others from saving them (P06). Interestingly, some teens acknowledged scenarios where they did not mind screenshots being taken, as these could serve as a form of validation, for instance, if something was \textit{``funny''} (P11) or captured special \textit{``moments''} (P18) with friends. Therefore, they proposed having more control over the screenshot function, such as a default block with the option to \textit{``allow screenshots''} when desired (P16). This approach would enable users to decide when to permit screenshots, balancing privacy with the desire to share certain content more freely.

Four participants (P01, P08, P11, P14) suggested the concept of \textit{conditional ephemerality} to address concerns about boundary violations in messaging. They appreciated Snapchat's feature of ephemeral messages with the option to save them and proposed similar functionalities for other messaging platforms. According to them, ephemerality provides a sense of comfort for \textit{``just in case... something happens''} (P14), yet they acknowledged situations where permanent messages are preferable. For instance, receiving a message \textit{``at 6 am''} and wanting to revisit it \textit{``at 9 am''} requires the messages to be available for reference and memory (P11). Additionally, they noted that in scenarios involving \textit{``issues or drama,''} including \textit{``legal issues,''} the ephemerality of messages complicates the process of tracing back conversations and can lead to challenges in proving claims (P08). Therefore, they proposed a more flexible messaging system that allows users to choose between ephemeral and permanent messages based on the context and necessity.

\vspace{2mm}
\noindent\textbf{Time collapse.} Ten participants (P02, P03, P06, P07, P13, P15-P19) raised concerns about digital spaces creating a \textit{``time collapse,''} where the past, present, and future merge~\cite{brandtzaeg2018time}. Participants noticed a shift in their privacy boundaries over time, making them uncomfortable with their earlier posts, such as \textit{``random posts''} (P06) and \textit{``every single thought''} (P13), which they no longer wish to be shared on their social media. Additionally, there was an apprehension about the long-term impact of their current online activities. For example, P13 expressed, \textit{``Digital footprints... I was taught that they will come back to haunt you and you see it nonstop. ... I don’t want the things I'm doing now as a teenager to haunt me in 10 years.''} P19 added, \textit{``anyone, including potential employers or universities, might see it in the future.''} These concerns were particularly challenging to address, as participants felt the need to anticipate judgment by a collapsed time of the future with potentially vastly different cultural or social norms. As P07 shared, \textit{``I might not like the way I look in two years or something like that. Or I might feel that that was kind of like a weird thing to post in a couple years.''}

Six participants (P01, P11, P14, P15-P17) expressed the importance of being able to easily \textit{revisit and reverse} content, such as through deleting or archiving posts. This preference stems from the fact that non-ephemeral posts \textit{``last longer and are more visible,''} leading to concerns that followers who join later can \textit{``go back and see what [they were] doing like two years ago''} (P17). While many participants actively engaged in revisiting and reversing their content, they also felt that the process could be more user-friendly. For instance, P17 pointed out the inconvenience of deleting comments on Instagram, describing the process as \textit{``really annoying.''} An additional suggestion from P16 was to receive a notification a few days after posting, asking \textit{``Do you want to delete this post? It's been two days. Do you want to delete this post, or are you comfortable with it staying here for an extended period of time?''} This feature would provide users with a timely reminder to reassess their comfort with the content's continued visibility.

Eleven participants (P02, P03, P05-P07, P09, P13, P14, P16-P18) also proposed the idea of content \textit{ephemerality}, viewing it as a form of \textit{``assurance that nobody's really scrutinizing it too much,''} given that \textit{``it's not something we can go back and look at later''} (P05). However, since ephemerality is already a feature in many platforms, and the participants generally expressed satisfaction with their experiences, we decided not to develop a prototype for this particular design idea in our design evaluation survey.

\subsubsection{Fear of Online Hostility}
Another frequently mentioned source of anxiety among the participants was the general unpredictability and perceived risks within social media environments, leaving them vulnerable and anxious as they navigated these platforms. To address these concerns, participants suggested enforcing robust community standards, enhancing security measures, and implementing features like pseudonymity to mitigate online hostility and cybersecurity breaches.

\vspace{2mm}
\noindent\textbf{Hostile environment.} Another source of fear that fourteen (P01-P08, P10, P12, P13, P15, P17, P18) of the participants mentioned was concerns about the chaotic or negative vibe of a specific platform, which often created a sense of discomfort or fear of privacy issues due to the overwhelming and often hostile environment.  P18, for example, voiced concerns about encountering \textit{``not very good influencers''} on Snapchat, contributing to a generally \textit{``not so positive''} vibe. Although the level of disorder on specific platforms may not be directly related to privacy concerns, it still prevented participants from feeling \textit{``safe''} (P18), thereby indirectly increasing their sense of vulnerability. This contrasts with P08's experience on Pinterest, where they felt \textit{``a lot safer''} due to the absence of \textit{``malicious''} users and an environment focused on \textit{``look[ing] at pretty aesthetic things.''} The teens' concerns were exacerbated by their perception that certain platforms do not prioritize user safety and fail to adequately regulate such disorders. P15's comment highlights this sentiment: \textit{``Twitter's a hellscape and it's awful and especially with the new leadership people get away with harassment really really easily.''}

Twelve participants (P01-P03, P05, P07, P09, P10, P13, P15, P17-P19) expressed a desire for more robust enforcement of \textit{community standards} on social media platforms. They observed considerable \textit{``risks''} (P18) and \textit{``negativity''} (P13) on major platforms like Instagram and Twitter and believed that mitigating these issues was crucial. One method they proposed was to regulate negativity by enabling users to report others and restricting toxic users. For instance, P02 suggested that \textit{``a third party could review and decide to remove a rude comment or prevent the user from posting for 30 days.''} P10 offered another approach, noting that negative comments often get posted on platforms like Instagram or Twitter without any hindrance. They advocated for a more preventative approach where such comments would be blocked from posting. P17 raised concerns about the complexity of privacy policies and the necessity for clearer enforcements, noting that \textit{``teenagers don't really read those because they're long and complicated.''} They suggested a simpler, more frequent reminder system, such as notifications when posting or commenting, to remind users to \textit{``make sure this is appropriate''} and to contribute to \textit{``keeping a good community.''}

\vspace{2mm}
\noindent\textbf{Cybersecurity breach.} Nine participants (P01, P02, P04-P06, P09, P13, P14, P17) expressed concerns about online threats that stem from unauthorized access or misuse of personal digital information. Much of this fear arose from indirect victimization experiences. P14 shared a specific instance: \textit{``A couple years ago... one of those online therapy websites that you pay for and the people got access to all the records and were able to like blackmail people. They were like if you don't give us money, we'll tell your relative you don't like them that kind of thing. And while I don't think it would be that extreme for a group of teenagers, sometimes there's just that chance of somebody that just for some reason didn't like you decided to get into your phone or your account and just like went and did bad things.''} Many participants mentioned various problems such as \textit{``data leak''} and \textit{``impersonation''} (P09), \textit{``doxxing''} and \textit{``bots''} (P06), and \textit{``child trafficking and... child pornography''} (P13). While they were concerned, the perception that these issues \textit{``happen to everyone''} made certain participants, like P05, feel \textit{``just not really comfortable''} while others felt these threats were beyond their control, resigning to the belief that \textit{``it just happens''} (P09).

Five participants (P02, P03, P09, P11, P15) suggested that \textit{enhanced security}, including options like \textit{pseudonymity}, could alleviate their cybersecurity breach concerns. They appreciated existing security measures, such as mandatory password changes when unusual login activities are detected and the use of two-factor authentication (P11). The ability to remain pseudonymous was particularly valued. P02 liked not having their account connected to a phone number, reducing stalker concerns. P09 preferred changing their username on different platforms for anonymity, and P15 appreciated the anonymity on Twitter, which allows interactions with a familiar circle while maintaining privacy from others.

\subsubsection{Fear of Personal Privacy Missteps}
Participants also expressed concerns about unknowingly engaging in actions on digital platforms that could lead to personal risks or unintended negative consequences. To address these concerns, participants suggested implementing prompts for self-regulation to prevent overexposure and clarifying privacy norms on platforms to help navigate and enforce proper privacy boundaries.

\vspace{2mm}
\noindent\textbf{Overexposure.} Six participants (P05, P08-P10, P13, P14) expressed worries about intentionally or unintentionally revealing too much personal information or aspects of their lives on digital platforms, leading to a loss of privacy and control over their digital footprint. The concern often revolved around `oversharing,' with some, like P08, unsure about \textit{``how much of [their] life is like appropriate to share.''} At times, this overexposure was facilitated by the affordances of social media platforms that make sharing easier. For instance, P05 pointed out that on BeReal, it is \textit{``concerning''} how \textit{``really easy to accidentally make your BeReal like public''} and inadvertently share their location. P14 shared a similar experience with Snapchat, where they \textit{``just send like random stuff on accident like even just like of [their] pocket or something it'll get bumped.''} These accidental shares sometimes resulted in disclosing information intended to be more private.

Six participants (P02, P05, P07, P08, P12, P17) suggested that \textit{prompts for self-regulation} could help them avoid overexposing or oversharing online. For instance, P12 noticed they were more cautious on Instagram than on Snapchat, as Instagram requires \textit{``clicking at least two or three buttons''} to post, encouraging more thoughtfulness. P07 proposed a feature that asks \textit{``Are you sure you want to post this?''} and displays the user's followers as a reminder. P08 agreed, saying such prompts provide a moment to reconsider: \textit{``Do I really want to share this? Will I regret this in the long run?''} This offers a chance to reassess one's confidence in what they're about to share. P12 also mentioned the idea of \textit{``daily posting limits''} to prevent \textit{``oversharing''}.

\vspace{2mm}
\noindent\textbf{Overstepping privacy norms.} Fifteen participants (P02, P03, P05-P09, P11-P14, P16-P19) expressed apprehension regarding the inadvertent overstepping of subtly established privacy expectations on social media platforms, which could potentially result in discomfort among other users or friends. Although these concerns are not privacy issues in themselves, they led to difficulties in navigating proper privacy management. For example, many participants felt pressured to add people against their will, as P03 mentioned, \textit{``pressured to add people [are] not that close to your account.''} They also found it challenging to remove friends they were uncomfortable with because, as P09 noted, \textit{``then you know you're like the aggressor in that situation.''} Furthermore, even when desiring privacy boundaries, participants found it awkward to enforce them, especially with friends. P02 shared that they found it  \textit{``awkward to say that you don't want things to be screenshotted,''} and P16 found it \textit{``awkward to set boundaries.''} These privacy norms were often implicit and varied across participants and their individual friends. However, most were highly vigilant about not overstepping any privacy norms, and the possibility of overstepping these norms and the resulting inability to enforce privacy due to fear of doing so were seen as difficult to avoid.

Thirteen participants (P01-P05, P07, P10, P11, P13-P17) emphasized the need for platforms to \textit{clarify privacy norms,} functioning as an \textit{``air cover.''} P02 underscored the importance of platforms stating explicitly that it is okay to reject friend requests, providing \textit{``reassurance''} in complex social situations. P13 recommended adding disclaimers such as, \textit{``Always make sure you know who you're following. You're always in control. You don't have to add someone if you don't want to,''} empowering users to make informed choices. P15 highlighted the need for platforms to clearly explain their privacy policies, thereby fostering a \textit{``culture''} that prioritizes user privacy and safety and encourages users \textit{``to set boundaries,''} as P03 stated. The participants observed that current platforms fail to communicate these norms effectively. P05 noted that people often inadvertently breach privacy because it is \textit{``assumed to be common knowledge.''} However, they remarked, \textit{``it's kind of been proven that that's not necessarily true,''} highlighting a discrepancy between assumed understanding and actual adherence to privacy norms. In essence, clarifying privacy norms and empowering users to use these guidelines as a rationale to prioritize their privacy, especially in situations where it might seem inappropriate or awkward, can be helpful.

\subsubsection{Vague and Persistent Fear} Many participants (P02-P04, P07-P10, P13-P16, P18, P19) often described experiencing non-specific, underlying feelings of threat or unease related to their general interactions on social media without being able to pinpoint a clear source. For instance, P03 mentioned, \textit{``I don't post my face right now, but I have an inner monologue that keeps telling me what I'm doing could get me into trouble. It's really just paranoia.''} Some expressed general concerns about social media usage, influenced by their parents' cautiousness. P04 noted that their parents are \textit{``very cautious''} about social media and advised them that they should \textit{``[not] get social media at all''} (P04). Media portrayals also fueled such fears. Despite perceiving that in the movies they are \textit{``overdramatized''} and \textit{``very exaggerated''} (P04), they still harbored significant concerns, particularly on public accounts where \textit{``there's a lot of weird stuff going on, like predators''} (P14). Since the sources of these fears were elusive, the teens experienced persistent anxiety. Even after implementing all known mitigation strategies and precautions, a vague and lingering anxiety remained, leaving them uncertain about how to alleviate this fear, suggesting the possibility of dysfunctional fear.

While the source of vague fear cannot be specified or directly mitigated, five participants (P07, P09, P13, P17, P18) shared design ideas that would provide them with \textit{reassurance} about their privacy fear in general. P13 proposed making it simpler for users to verify their privacy settings, suggesting, \textit{``So after I post, if I'm unsure whether I included someone in the privacy settings, I should be able to check it quickly, in just two or three clicks, rather than navigating through settings. It should be more straightforward.''} P17 emphasized the importance of control over one's profile, \textit{``Especially with your own profile, having a lot of control over what people post and how you express yourself''} contributes to this reassurance. P18 agreed, noting the benefits of options like \textit{``closing my comments or likes''} to make things \textit{``more private for yourself,''} which they found to be a \textit{``huge reassurance.''} Additionally, P13 suggested that platforms could enhance feelings of safety by sharing positive affirmations, such as \textit{``you are safe, you are loved, and we care about you.''}

\input{tables/prototypes}
\input{tables/features}

\subsection{How Does Design Affect Teens' Privacy-related Fears?}
We developed ten prototypes corresponding to each of the ten design ideas suggested by teens during our co-design interviews. Although there are various ways to implement the specific features of each design idea, we chose to use prototypes as design probes instead of conceptual design ideas. To elicit more concrete feedback from teens, we presented specific implementations of our design ideas. While these examples represent just a few possible approaches, they facilitated clearer understanding and discussion. The core rationale behind our designs was to empower users with enhanced privacy control, offering them more choices and greater autonomy than existing features on mainstream platforms or those explored in prior research. Our aim was to ensure users feel empowered and in control of their privacy, thereby addressing dysfunctional worries. Importantly, we focused on implementing actual protective measures rather than creating privacy theater, thus providing substantive safeguards while alleviating vague and/or lingering anxiety. A comprehensive list of design strategies, sample prototypes, and detailed descriptions provided to participants during the follow-up design evaluation survey are available in Table \ref{tab:features}. Additionally, example screenshots of all prototypes are displayed in Figure \ref{tab:prototypes}\footnote{The license plate number shown in (h) is a fictional placeholder, not an actual vehicle registration.}.

Overall, teens showed high interest in the prototypes we shared throughout the design evaluation survey. As illustrated in Figure \ref{tab:individual_evaluations}, follow-up survey respondents expressed interest in and positive reactions to all features. While some prototypes were preferred more than others, one-sample t-test results revealed that the mean responses to the questions `My reaction to this feature is,' `I would be interested in trying this feature on my [public/private] account,' `With this feature, I would feel less worried about privacy-related problems on my [public/private] account,' and `With this feature, the likelihood of privacy-related problems on my [public/private] account will decrease,' were significantly higher than 3 (`Neutral' and `Neither agree nor disagree,' respectively) for all ten features. Conversely, one-sample t-test results showed that the mean responses to the questions `This feature might make my overall experience using my [public/private] account worse,' `I find this feature to be annoying,' and `If I use this feature on my [public/private] account, I'm concerned it might lead to awkward or uncomfortable situations with my friends or people I know in real life' were all significantly lower than 3 for all features. \footnote{Despite the hypothetical nature of several questions, participants consistently rated their confidence in their responses to the question `How confident are you about the responses you gave above?' as nearly 5 (`Very confident'), which is significantly higher the neutral rating (`Moderately confident').} 

Furthermore, as seen in Figure \ref{tab:overall_evaluations}, with the features applied together, it would decrease their fear in all the different types of fear we identified above. This revealed that the prototypes were generally perceived as possibly effective in addressing dysfunctional fear, as they decreased the perceived risk and the actual risk while not revealing many new tradeoffs, such as causing discomfort or potential problems with other users. Still, through free response questions, they shared some of their concerns and areas for improvement, which we explore below. Given the overwhelmingly positive reactions to the Likert-scale questions, \textit{we deliberately focused on examining the potential pitfalls of these features} to provide a balanced perspective and highlight possible trade-offs or concerns associated with implementing these privacy measures.

\vspace{2mm}
\noindent\textbf{Account protection.} As an example of \textit{boundary/visibility control}, we designed a feature (Fig. \ref{tab:prototypes}(a)) to address issues with hostile interactions from strangers while balancing users' concerns about limiting their audience against the opportunity to reach a larger network. This feature allows users to experiment with and choose their own balance in this trade-off by filtering other users based on the number of warnings they've received on the platform. This empowers privacy-conscious users to implement stricter audience limitations, while those prioritizing network expansion can minimize restrictions.

Participants responded very positively to this feature, as evidenced by the following results:
\begin{itemize}
    \item Overall positive reaction: \((mean=4.27, sd=0.890, t(135)=16.7, p<.0001, d=1.43)\)
    \item Interest in trying the feature: \((mean=4.09, sd=0.954, t(135)=13.3, p<.0001, d=1.14)\)
    \item Belief that the feature would reduce privacy worries: \((mean=4.09, sd=1.00, t(135)=12.7, p<.0001, d=1.09)\)
    \item Expectation of decreased privacy concerns: \((mean=3.91, sd=1.04, t(135)=10.2, p<.0001, d=0.874)\)
\end{itemize}

Importantly, participants also:
\begin{itemize}
    \item Strongly disagreed that the feature would worsen their overall social media experience: \((mean=1.92, sd=1.01, t(135)=-12.5, p<.0001, d=-1.07)\)
    \item Did not find the feature annoying: \((mean=1.84, sd=0.929, t(135)=-14.6, p<.0001, d=-1.25)\)
    \item Moderately disagreed that the feature would cause awkward situations with friends: \((mean=2.22, sd=1.18, t(135)=-7.72, p<.0001, d=-0.662)\)
\end{itemize}

Specifically, participants acknowledged its potential as \textit{``a good step in between private and fully public''} (R056) and its empowerment of users by granting them enhanced \textit{``control''} (R124) and \textit{``security''} (R118). However, feedback also included concerns about the accuracy and transparency of the red flag system. Suggestions were raised for clearer definitions and criteria for red flagging, as well as a more nuanced system where flags could expire or become \textit{``gray flags''} (R005) over time. Questions were raised about the practicality of quantifying red flags, accompanied by requests for manual override options to maintain connections despite red flags. Private account users noted potential redundancy with their existing selective audience management. Additionally, concerns were expressed regarding the possibility of users circumventing the system by creating new accounts.

\vspace{2mm}
\noindent\textbf{Categorized viewer list.} Our \textit{interaction transparency} feature (Fig. \ref{tab:prototypes}(b)) builds upon existing functionalities like Instagram's Story viewer list but addresses a key limitation identified by participants: lengthy user lists often fail to provide complete audience transparency~\cite{DeVito2017-by} or awareness of potentially unwanted profile interactions. To enhance this, we categorized viewers based on the closeness of their interactions with the user. This approach helps users identify unwanted interactions without compromising the positive feedback they receive from seeing that close contacts view their profiles.

Participants responded very positively to this feature and expressed a moderately high level of confidence in its effectiveness for enhancing privacy:
\begin{itemize}
    \item Overall positive reaction: \((mean=4.10, sd=1.03, t(135)=12.4, p<.0001, d=1.07)\)
    \item Interest in trying the feature: \((mean=3.92, sd=1.22, t(135)=8.76, p<.0001, d=0.751)\)
    \item Belief that the feature would reduce privacy worries: \((mean=3.68, sd=1.06, t(135)=7.53, p<.0001, d=0.646)\)
    \item Expectation of decreased privacy concerns: \((mean=3.54, sd=1.07, t(135)=5.83, p<.0001, d=0.500)\)
\end{itemize}

Moreover, participants also:
\begin{itemize}
    \item Strongly disagreed that the feature would worsen their overall social media experience: \((mean=1.92, sd=1.01, t(135)=-12.5, p<.0001, d=-1.07)\)
    \item Did not find the feature annoying: \((mean=1.84, sd=0.929, t(135)=-14.6, p<.0001, d=-1.25)\)
    \item Moderately disagreed that the feature would cause awkward situations with friends: \((mean=2.22, sd=1.18, t(135)=-7.72, p<.0001, d=-0.662)\)
\end{itemize}

The feature was well-received, primarily because it enables users to more easily detect an \textit{``unwanted person''} (R048) viewing their account. By organizing viewers into categories, users can quickly scan for anomalies or unexpected interactions, enhancing their sense of control and awareness. Nevertheless, concerns were expressed that it might become \textit{``almost too intricate of a feature to enjoy the fun in posting even if the intention is to protect privacy''} (R013) and that it would make it \textit{``too easy to get caught up in looking at who viewed your posts''} (R020). Suggestions to enhance the feature included sharing additional information, such as the \textit{``day and time last interacted''} (R046). Participants also proposed an option to toggle the feature on and off, acknowledging that this level of transparency might be uncomfortable for some users.

\vspace{2mm}
\noindent\textbf{Auto-delete with save.} This feature (Fig. \ref{tab:prototypes}(c)) was designed for \textit{conditional ephemerality}, providing automatic deletion while giving users the option to save posts. It also alerts other users when conversations are saved, addressing potential privacy concerns when the default ephemerality is overridden. This feature elicited the most diverse responses among participants. Many appreciated its ability to \textit{prevent people from bringing up old things''} (R067), particularly in protecting privacy from \textit{strict parents''} (R069). The notification system for saved conversations without \textit{``asking for consent''} (R016) was particularly valued.

However, concerns were raised about losing the benefits of permanent content, as it can be \textit{fun looking back at old convos''} (R006), or it might make retrieving information from the chat difficult (R018). Feedback highlighted issues such as potential \textit{bullying''} (R125) and the lack of \textit{evidence from the chat''} when needed (R115). Suggestions included options to \textit{turn it on or off''} (R077), or the ability to \textit{``never delete chats with friends [they] talk to often''} (R124).

Quantitative data revealed generally positive responses:
\begin{itemize}
    \item Moderately positive reaction: \((mean=3.43, sd=1.34, t(135)=3.77, p<.001, d=0.323)\)
    \item Moderate interest in trying the feature: \((mean=3.40, sd=1.40, t(135)=3.37, p<.001, d=0.289)\)
    \item Moderate belief that the feature might reduce privacy worries: \((mean=3.43, sd=1.25, t(130)=3.92, p<.001, d=0.343)\)
    \item Moderate expectation of decreased privacy concerns: \((mean=3.32, sd=1.22, t(135)=3.03, p=.003, d=0.260)\)
\end{itemize}

Moreover, participants also:
\begin{itemize}
    \item Moderately disagreed that the feature would worsen their overall social media experience: \((mean=2.49, sd=1.27, t(135)=-4.72, p<.0001, d=-0.405)\)
    \item Felt the feature is not too annoying: \((mean=2.75, sd=1.39, t(135)=-2.10, p=.037, d=-0.180)\)
    \item Did not strongly agree that the feature would cause awkward situations with friends: \((mean=2.76, sd=1.36, t(135)=-2.08, p=.040, d=-0.178)\)
\end{itemize}

\vspace{2mm}
\noindent\textbf{On-demand screenshot blocking.} This feature (Fig. \ref{tab:prototypes}(d)) addresses teens' complex relationship with \textit{screenshot control}. Our participants expressed nuanced views on screenshots, informed by their experiences with existing apps like Snapchat that notify users when screenshots are taken. They acknowledged that screenshots could threaten privacy by enabling out-of-context sharing but also recognized legitimate reasons for taking them, such as preserving positive content. Participants noted that accidental screenshots can lead to awkward situations due to automatic notifications, and some preferred not to know when screenshots are taken to avoid overthinking. Given this range of preferences, we designed a feature allowing users to disable screenshots for individual posts. This approach aims to provide flexible control over content sharing, shape platform norms by conveying that screenshots are not always acceptable, and respect the original poster's intentions.

As expected, this feature was appreciated for its ability to block screenshots and for offering the option to apply this \textit{``not to your whole profile, but just per post''} (R069). Concerns were raised about its effectiveness, noting potential circumvention by users who might use \textit{``a second device or find a way to screenshot/screen record anyway, as many do with Snapchat''} (R078). Feedback included worries about potential exploitation for malicious purposes, with the possibility for \textit{``some people to use this feature to post cruel or harmful things without risk of them being saved and/or shared''} (R029). While addressing such behind-the-scenes betrayal directly through social media design is challenging, this feedback highlights the importance of clarifying privacy norms so that users feel empowered to defend themselves when necessary.

Participants perceived this feature very positively and highly effective in addressing privacy concerns:
\begin{itemize}
    \item Overall positive reaction: \((mean=4.41, sd=0.946, t(135)=17.4, p<.0001, d=1.49)\)
    \item Strong interest in trying the feature: \((mean=4.21, sd=1.08, t(135)=13.1, p<.0001, d=1.12)\)
    \item Strong belief that the feature might reduce privacy worries: \((mean=4.32, sd=0.948, t(135)=16.2, p<.0001, d=1.39)\)
    \item High expectation of decreased privacy concerns: \((mean=4.17, sd=0.931, t(135)=14.6, p<.0001, d=1.26)\)
\end{itemize}

Participants also indicated minimal drawbacks or trade-offs:
\begin{itemize}
    \item They strongly disagreed that the feature would worsen their overall social media experience: \((mean=1.85, sd=1.08, t(135)=-12.4, p<.0001, d=-1.06)\)
    \item They did not find the feature annoying: \((mean=1.93, sd=1.20, t(135)=-10.5, p<.0001, d=-0.897)\)
    \item They disagreed that the feature would cause awkward situations with friends: \((mean=2.38, sd=1.29, t(135)=-5.59, p<.0001, d=-0.479)\)
\end{itemize}

\input{tables/individual_evaluations}

\input{tables/overall_evaluations}

\vspace{2mm}
\noindent\textbf{Reminder to review.} This feature (Fig. \ref{tab:prototypes}(e)) facilitates \textit{revisit and reverse}, addressing the dynamic nature of teens' identities and evolving social norms. It prompts users to reassess past content at different time intervals, recognizing that teens' perspectives on their posts may change over time. By sending notifications, the feature eliminates the need for manual searching and reduces the likelihood of overlooking old content. This approach empowers users to make informed decisions about keeping or modifying past posts, encouraging regular reflection on their digital footprint while balancing the preservation of meaningful memories with the curation of their current online presence.

The feature was well-received by participants for helping users to \textit{``stay current with [their] beliefs and values''} (R039) by allowing them to \textit{``reconsider or delete posts that reflect `the old me'''} (R025). It was praised for providing an opportunity for \textit{``self-reflection''} and \textit{``reassurance''} (R132). However, there were concerns about the potential for notifications to become \textit{``annoying''} (R018). Suggestions for improvement included providing \textit{``tips on what should stay up and what should not''} (R002) and adding an \textit{``archive option alongside a delete option''} (R026). A participant who identified themself as an \textit{``infrequent poster''} (R030) felt the feature was unnecessary, as they are likely aware of their social media content. Additionally, there was worry that the feature might inadvertently remind users of \textit{``things made in the past that you may not be comfortable with''} (R099). Again, a desire for the ability to \textit{``toggle this on and off''} (R030) was highlighted.

The reactions were confirmed in our quantitative data:
\begin{itemize}
    \item Overall positive reaction: \((mean=4.13, sd=1.01, t(135)=13.0, p<.0001, d=1.12)\)
    \item Strong interest in trying the feature: \((mean=3.92, sd=1.19, t(135)=9.03, p<.0001, d=0.775)\)
    \item Moderate belief that the feature might reduce privacy worries: \((mean=3.68, sd=1.15, t(135)=6.84, p<.0001, d=0.586)\)
    \item Moderate expectation of decreased privacy concerns: \((mean=3.59, sd=1.08, t(135)=6.36, p<.0001, d=0.546)\)
\end{itemize}

Participants also reported minimal drawbacks:
\begin{itemize}
    \item Disagreement that the feature would worsen their social media experience: \((mean=1.83, sd=0.931, t(135)=-14.6, p<.0001, d=-1.26)\)
    \item Low perception of the feature as annoying: \((mean=2.06, sd=1.12, t(135)=-9.79, p<.0001, d=-0.840)\)
    \item Disagreement that the feature would cause awkward situations with friends: \((mean=1.85, sd=0.942, t(135)=-14.3, p<.0001, d=-1.23)\)
\end{itemize}

\vspace{2mm}
\noindent\textbf{Red flag and follow-up.} This feature (Fig. \ref{tab:prototypes}(f)) was designed to assist with the enforcement of \textit{community standards}, addressing participants' skepticism about mainstream social media apps' commitment to user privacy. The feature was appreciated for providing follow-up, which gives users \textit{``peace of mind''} (R026) as intended. However, concerns were raised about the validity and feasibility of the ``red flag'' feature. As R045 pointed out, \textit{``If this [hypothetical] platform becomes successful, you need to realize the number of people on the platform. The report system would get overwhelmed and not do its job. Even with an AI moderator, so many people would get wrongfully flagged. Additionally, it's so easy nowadays to get something like a fake email address or a VOIP phone number, making it appear as if different people are reporting. Even if you based it on IP address, VPNs and alternate devices exist.''} Other worries included the potential for the feature to be \textit{``abused''} (R102), with people flagging \textit{``just to be annoying''} (R038). While detailing the operation of the red flag system is beyond the scope of this study, the feedback highlights the importance of the safety reassurance that the system provides users.

Despite the concerns, the feature was very well received, as corroborated in our quantitative data:
\begin{itemize}
    \item Overall positive reaction: \((mean=4.45, sd=0.806, t(135)=21.0, p<.0001, d=1.80)\)
    \item High interest in trying the feature: \((mean=4.11, sd=0.979, t(135)=13.2, p<.0001, d=1.13)\)
    \item Strong belief that the feature might reduce privacy worries: \((mean=4.01, sd=0.947, t(135)=12.4, p<.0001, d=1.06)\)
    \item Expectation of decreased privacy concerns: \((mean=3.85, sd=0.995, t(135)=9.91, p<.0001, d=0.850)\)
\end{itemize}

Participants also reported minimal drawbacks:
\begin{itemize}
    \item Strong disagreement that the feature would worsen their social media experience: \((mean=1.79, sd=0.977, t(135)=-14.5, p<.0001, d=-1.24)\)
    \item Very low perception of the feature as annoying: \((mean=1.84, sd=1.01, t(135)=-13.4, p<.0001, d=-1.15)\)
    \item Disagreement that the feature would cause awkward situations with friends: \((mean=2.10, sd=1.12, t(135)=-9.31, p<.0001, d=-0.798)\)
\end{itemize}

\vspace{2mm}
\noindent\textbf{Pseudonymity mode.} This feature (Fig. \ref{tab:prototypes}(g)) aims to \textit{enhance security} by allowing users to detach their real-world identity from shared content while maintaining social connectivity. The pseudonymity mode can be selectively applied to non-followers, users not followed, or specific individuals. This granular control enables teens to balance privacy protection with social connections.

This feature received mixed reactions. It was praised for enabling users to \textit{``manage your social media presence completely without having to block other users''} (R101), offering an additional \textit{``layer of protection,''} (R039), especially for public posts. However, there were concerns that it might encourage people to \textit{``express harmful opinions more comfortably''} (R003), and some felt that \textit{``just blocking the user''} might be a simpler solution (R108). Additionally, participants brought to attention that the feature could be \textit{``used maliciously, such as by anonymous users viewing others' posts without being identified''} (R014). There were also worries that this mode might make it \textit{``harder to gain more followers''} (R002) as it is \textit{``harder for someone to find you''} (R050). Concerns were also expressed about whether the term \textit{``pseudonymity''} would be familiar to teens. Moreover, reluctance towards pseudonymous communication was noted, with sentiments like \textit{``you shouldn't be on social media if you are trying to hide your identity''} (R044). This feedback highlights that privacy features can sometimes exacerbate privacy concerns and that users' preferences and views vary greatly, emphasizing the need for having the features be optional.

The mixed reactions to this feature were reflected in our survey data:
\begin{itemize}
    \item Overall positive reaction: \((mean=3.80, sd=1.23, t(135)=7.61, p<.0001, d=0.653)\)
    \item Interest in trying the feature: \((mean=3.64, sd=1.30, t(135)=5.72, p<.0001, d=0.491)\)
    \item Belief that the feature might reduce privacy worries: \((mean=3.83, sd=1.16, t(135)=8.37, p<.0001, d=0.717)\)
    \item Expectation of decreased privacy concerns: \((mean=3.76, sd=1.19, t(135)=7.43, p<.0001, d=0.637)\)
\end{itemize}

Despite concerns about potential drawbacks, participants generally did not anticipate significant negative impacts:
\begin{itemize}
    \item Disagreement that the feature would worsen their social media experience: \((mean=2.29, sd=1.16, t(135)=-7.09, p<.0001, d=-0.608)\)
    \item Low perception of the feature as annoying: \((mean=2.23, sd=1.21, t(135)=-7.44, p<.0001, d=-0.638)\)
    \item Mild disagreement that the feature would cause awkward situations with friends: \((mean=2.49, sd=1.27, t(135)=-4.66, p<.0001, d=-0.399)\)
\end{itemize}

\vspace{2mm}
\noindent\textbf{Personal information alert.} This feature (Fig. \ref{tab:prototypes}(h)), designed to nudge for self-regulation, was recognized for helping users avoid \textit{``overlook[ing] details they may have missed''} (R032), particularly beneficial for those who \textit{``post more impulsively''} (R070). Concerns were raised about its accuracy and potential intrusiveness. Some users were apprehensive that it could \textit{``sense things wrong and be annoying''} (R052), and there were privacy concerns regarding the perception of it \textit{``checking what [they] post''} (R014). Suggestions for implementation included having the option to turn it \textit{``on and off, but with the default setting as on''} (R011). Additionally, users thought the feature could be enhanced by allowing in-app content blurring. This feedback suggests that sometimes, a privacy issue associated with a privacy feature is overlooked because the benefit of the feature is significantly larger and more widely accepted.

This feature received the most positive reactions from teens:
\begin{itemize}
    \item Overwhelmingly positive reaction: \((mean=4.64, sd=0.685, t(135)=27.9, p<.0001, d=2.39)\)
    \item Strong interest in trying the feature: \((mean=4.43, sd=0.814, t(135)=20.5, p<.0001, d=1.76)\)
    \item Strong belief that the feature might reduce privacy worries: \((mean=4.44, sd=0.796, t(135)=21.1, p<.0001, d=1.81)\)
    \item Strong expectation of decreased privacy concerns: \((mean=4.32, sd=0.900, t(135)=17.1, p<.0001, d=1.46)\)
\end{itemize}

Participants also strongly disagreed with potential drawbacks:
\begin{itemize}
    \item Strong disagreement that the feature would worsen their social media experience: \((mean=1.70, sd=1.03, t(135)=-14.8, p<.0001, d=-1.27)\)
    \item Very low perception of the feature as annoying: \((mean=1.71, sd=0.928, t(135)=-16.3, p<.0001, d=-1.39)\)
    \item Strong disagreement that the feature would cause awkward situations with friends: \((mean=1.46, sd=0.816, t(135)=-22.0, p<.0001, d=-1.88)\)
\end{itemize}

\vspace{2mm}
\noindent\textbf{User privacy norms.} Intended to \textit{clarify privacy norms}, this feature (Fig. \ref{tab:prototypes}(i)) was positively received in general. Participants appreciated its guidance in online safety management, emphasizing that \textit{``You aren't left to figure out how to protect your safety on your own''} (R041) and noting its role in \textit{``help[ing] establish norms for the platform''} (R060). It was commended for making \textit{``being cautious''}, not something to be \textit{``ashamed or embarrassed of''} (R026) and for ensuring \textit{``being respectful [is] normal''} (R047). Additionally, participants noted that with this feature, \textit{``users won't be able to deny knowing the rules and policies''} (R033). The feature's effectiveness in setting clear expectations was highlighted, described as \textit{``short''} (R023) and able to \textit{``very clearly but simply explain things''} (R032). However, some participants desired more specificity, wanting guidelines that are \textit{``more specific to potential scenarios''} and offer \textit{``more details of what constitutes `safe'''} (R050). Concerns were raised that some people might \textit{``just click off and not read or care about it''}; including \textit{``a quiz of some sorts''} was suggested to confirm users' understanding and ensure they \textit{``actually read over the guidelines''} (R035).

This feature was very well-received, with participants viewing it as highly necessary. However, they were less confident in its ability to address privacy concerns directly, highlighting the importance of shifting overall privacy norms:
\begin{itemize}
    \item Overall positive reaction: \((mean=4.20, sd=0.987, t(135)=14.2, p<.0001, d=1.21)\)
    \item Strong interest in trying the feature: \((mean=3.93, sd=1.03, t(135)=10.6, p<.0001, d=0.909)\)
    \item Moderate belief that the feature might reduce privacy worries: \((mean=3.59, sd=1.09, t(135)=6.28, p<.0001, d=0.539)\)
    \item Moderate expectation of decreased privacy concerns: \((mean=3.54, sd=1.11, t(135)=5.61, p<.0001, d=0.481)\)
\end{itemize}

Participants also strongly felt the feature would have minimal negative impacts:
\begin{itemize}
    \item Strong disagreement that the feature would worsen their social media experience: \((mean=1.82, sd=0.918, t(135)=-14.9, p<.0001, d=-1.28)\)
    \item Low perception of the feature as annoying: \((mean=2.01, sd=1.13, t(135)=-10.2, p<.0001, d=-0.877)\)
    \item Strong disagreement that the feature would cause awkward situations with friends: \((mean=1.70, sd=0.855, t(135)=-17.8, p<.0001, d=-1.52)\)
\end{itemize}

\vspace{2mm}
\noindent\textbf{``View as''.} Participants generally appreciated the feature (Fig. \ref{tab:prototypes}(j)) designed to enhance \textit{safety reassurance}. Despite its similarity to Facebook's ``View As'' function, we included it in our survey as most teens reported being unfamiliar with Facebook during interviews. They found it particularly useful \textit{``if you have certain things blocked from certain users''} (R012). The feature was perceived to be useful in allowing users to \textit{``step' into the shoes of others so you can make sure that what you're changing about your account is what you want others to see''} (R062), eliminating \textit{``the hassle of having to use different accounts to see what my main account looks like to others,''} which can be \textit{``very tiring''} (R039). However, there were views that this feature might be redundant for private accounts as \textit{``there aren't THAT many people too concerned with their private content''} (R044), and some felt that \textit{``on Instagram, your profile already looks pretty much how others will see it''} (R087). Participant reactions to the feature underscored the importance of psychological reassurance regarding privacy and safety for some users, even if the feature does not directly provide additional control.

Respondents showed a strong positive reaction to this feature:
\begin{itemize}
    \item Highly positive reaction: \((mean=4.40, sd=0.898, t(135)=18.2, p<.0001, d=1.56)\)
    \item Strong interest in trying the feature: \((mean=4.22, sd=0.956, t(135)=14.9, p<.0001, d=1.28)\)
    \item Belief that the feature might reduce privacy worries: \((mean=3.97, sd=1.03, t(135)=11.0, p<.0001, d=0.947)\)
    \item Expectation of decreased privacy concerns: \((mean=3.76, sd=1.06, t(135)=8.30, p<.0001, d=0.711)\)
\end{itemize}

Participants also strongly disagreed with potential negative impacts:
\begin{itemize}
    \item Strong disagreement that the feature would worsen their social media experience: \((mean=1.78, sd=0.932, t(135)=-15.3, p<.0001, d=-1.31)\)
    \item Very low perception of the feature as annoying: \((mean=1.75, sd=0.964, t(135)=-15.1, p<.0001, d=-1.30)\)
    \item Disagreement that the feature would cause awkward situations with friends: \((mean=1.90, sd=1.11, t(135)=-11.5, p<.0001, d=-0.989)\)
\end{itemize}

%% file: tables/prototypes.tex
\begin{figure}[!t]
    \centering
    \includegraphics[width=1.0\linewidth]{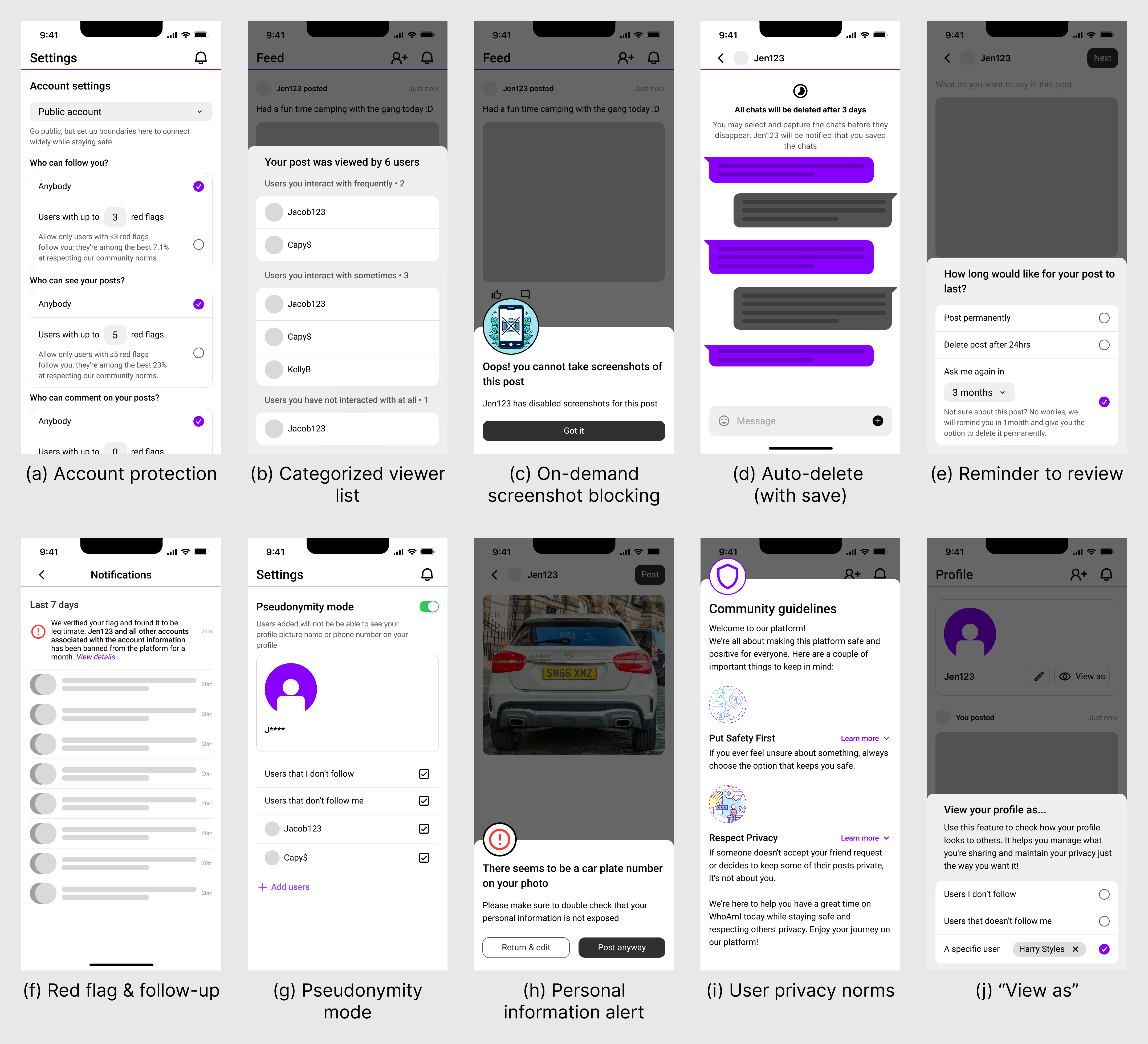}
    \caption{Sample screens of prototypes of the ten design approaches derived from the co-design study with teen participants. Survey respondents were provided additional screens for clarity. A complete taxonomy of the designs is available in Table \ref{tab:features}.}
    \Description {Figure 1 of ten screenshots labeled from (a) to (j), each depicting different user interface features from a social media application focused on privacy: (a) Account protection: A settings menu allowing users to control who can follow them and comment on their posts, with options to limit interactions based on a "red flag" system. (b) Categorized viewer list: A feed section showing a post with a list of viewers categorized by the frequency of interaction. (c) On-demand screenshot blocking: A notification alerting the user that screenshots cannot be taken of a particular post because the poster has disabled this function. (d) Auto-delete (with save): A message interface with an option for the user to set their chats to auto-delete after a certain period while still allowing them to save specific messages. (e) Reminder to review: A feature offering users the option to decide the duration of their post's visibility, ranging from 24 hours to permanent, with reminders to review this setting. (f) Red flag & follow-up: A notifications tab showing a red flag alert indicating that a post has been flagged and found to be in violation of community norms, with an option for more details. (g) Pseudonymity mode: A settings menu where users can activate a mode that hides their username and phone number on their public profile. (h) Personal information alert: An alert overlay on a photo within a post, warning the user that there appears to be a car plate number visible and advising them to ensure no personal information is exposed. (i) User privacy norms: A community guidelines page highlighting the importance of respecting privacy and offering tips for safe engagement on the platform, with specific reference to the respect of personal space and keeping user information private. (j) "View as": A profile customization feature that allows users to preview their profile as it appears to other users, with options to see how it looks to users they follow and those they don't, ensuring control over their public persona. Each screenshot is designed with a user-friendly interface, displaying clear text and interactive elements such as switches, dropdown menus, and informative pop-ups. These features collectively aim to enhance user privacy and provide personalized control over content sharing and interaction within the social media application.}
    \label{tab:prototypes}
\end{figure}

%% file: tables/features.tex
\begin{table*}[!t]
\small
\caption{A taxonomy of design ideas generated in co-design study. The example feature name and description are shown to the teens during the design evaluation survey.}
\label{tab:features}
    \begin{tabularx}{\textwidth}{@{}>{\raggedright\arraybackslash}p{4cm} >{\raggedright\arraybackslash}p{4cm} >{\raggedright\arraybackslash}X@{}}
    \toprule
    \textbf{Design Approach} & \textbf{Example Feature} & \textbf{Feature Description} \\
    \midrule
    Boundary control & Account protection (Fig. \ref{tab:prototypes}(a)) & ``Provides extra security layers for your public profile, keeping your info safe while you connect with more people.'' \\
    Interaction transparency & Categorized viewer list (Fig. \ref{tab:prototypes}(b)) & ``Sorts your audience based on how often you interact, so you can decide if less frequent contacts should stay on your follower list.'' \\
    (Conditional) Ephemerality & Auto-delete (with save) (Fig. \ref{tab:prototypes}(c)) & ``Automatically erases your content after a set time, but gives you a heads-up to save them if you want to keep them.'' \\
    Context control & On-demand screenshot blocking (Fig. \ref{tab:prototypes}(d)) & ``Stops others from taking screenshots of your posts if you want to, keeping your shared moments private.'' \\
    Revisit/reverse & Reminder to review (Fig. \ref{tab:prototypes}(e)) & ``Reminds you to look back at old posts and reassess if they're still appropriate or reflect who you are today.'' \\
    Community standards & Red flag \& follow-up (Fig. \ref{tab:prototypes}(f)) & ``Reports breaches of community rules and get notified when action is taken, ensuring your concerns are addressed.'' \\
    Enhanced security & Pseudonymity mode (Fig. \ref{tab:prototypes}(g)) & ``Completely disconnect your real identity from your account, making it untraceable to certain people you choose, for ultimate privacy.'' \\
    Prompts for self-regulation & Personal information alert (Fig. \ref{tab:prototypes}(h)) & ``Alerts you if you're about to share personal info, helping you think twice about your privacy.'' \\
    Clarification of privacy norms & User privacy norms (Fig. \ref{tab:prototypes}(i)) & ``Guides for safe online sharing, reminding you that being privacy-conscious is smart, not odd.'' \\
    Safety reassurance & ``View as'' (Fig. \ref{tab:prototypes}(j)) & ``Lets you see your profile through someone else's eyes, so you can be sure you're sharing only what you intend to.'' \\
    \bottomrule
    \end{tabularx}
\end{table*}

%% file: tables/individual_evaluations.tex
\begin{figure}
  \centering
  \begin{subfigure}{1.0\textwidth}
    \centering
    \includegraphics[width=0.9\linewidth]{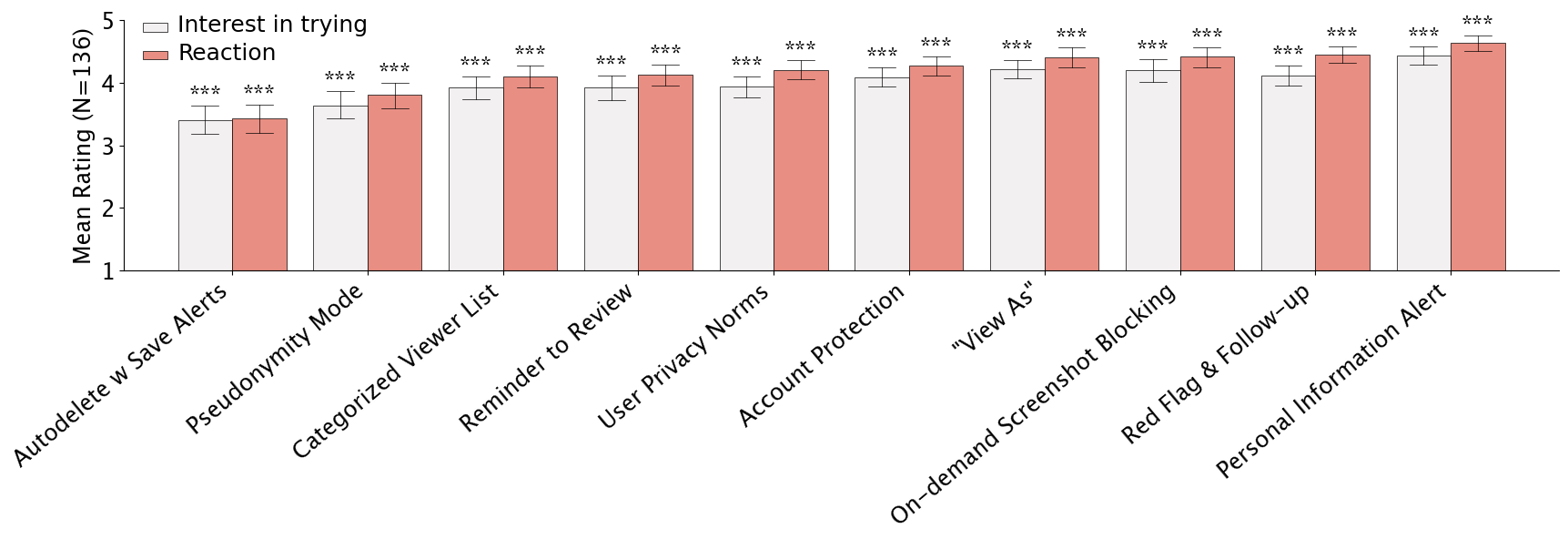}
    \caption{Interest and reaction toward features}
    \Description{Figure 2. (a) has a bar graph representing the mean rating of participants' interest in trying features (light red) and their reaction to them (dark red). Each feature is rated between 3 to 5 on a Likert scale, with higher values indicating a stronger positive response. Features include Autodelete, Save Alerts, Pseudonymity Mode, Categorized Viewer List, Reminder to Review, User Privacy Norms, View As, Account Protection, On-demand Screening/Blocking, Red Flag & Follow-up, and Personal Information Alert. All features show a high level of interest and positive reaction, with several features scoring close to 5.}
  \end{subfigure}
  \begin{subfigure}{1.0\textwidth}
    \centering
    \includegraphics[width=0.9\linewidth]{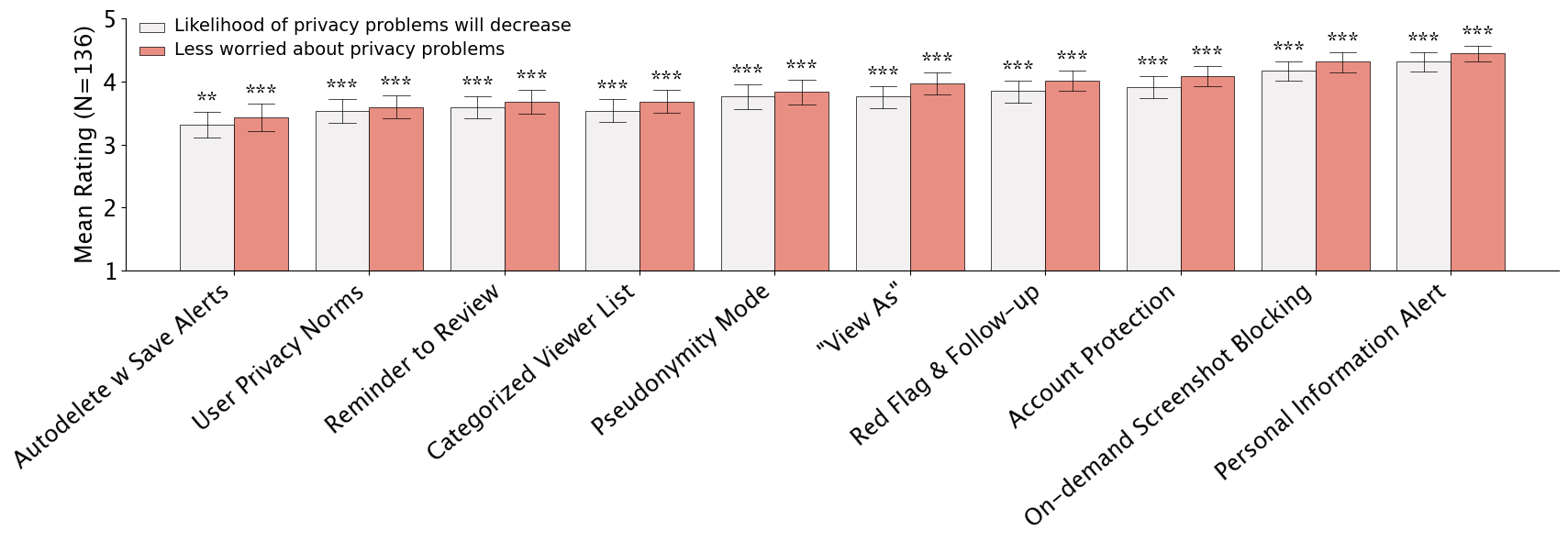}
    \caption{Privacy-related effectiveness of the features}
    \Description{Figure 2. (b) has a bar graph showing the mean rating of participants' views on the likelihood of privacy problems decreasing (light red) and being less worried about privacy problems (dark red) due to these features. All features are rated between 3 to 5, suggesting participants perceive them as effective in enhancing privacy}
  \end{subfigure}
  \begin{subfigure}{1.0\textwidth}
    \centering
    \includegraphics[width=0.9\linewidth]{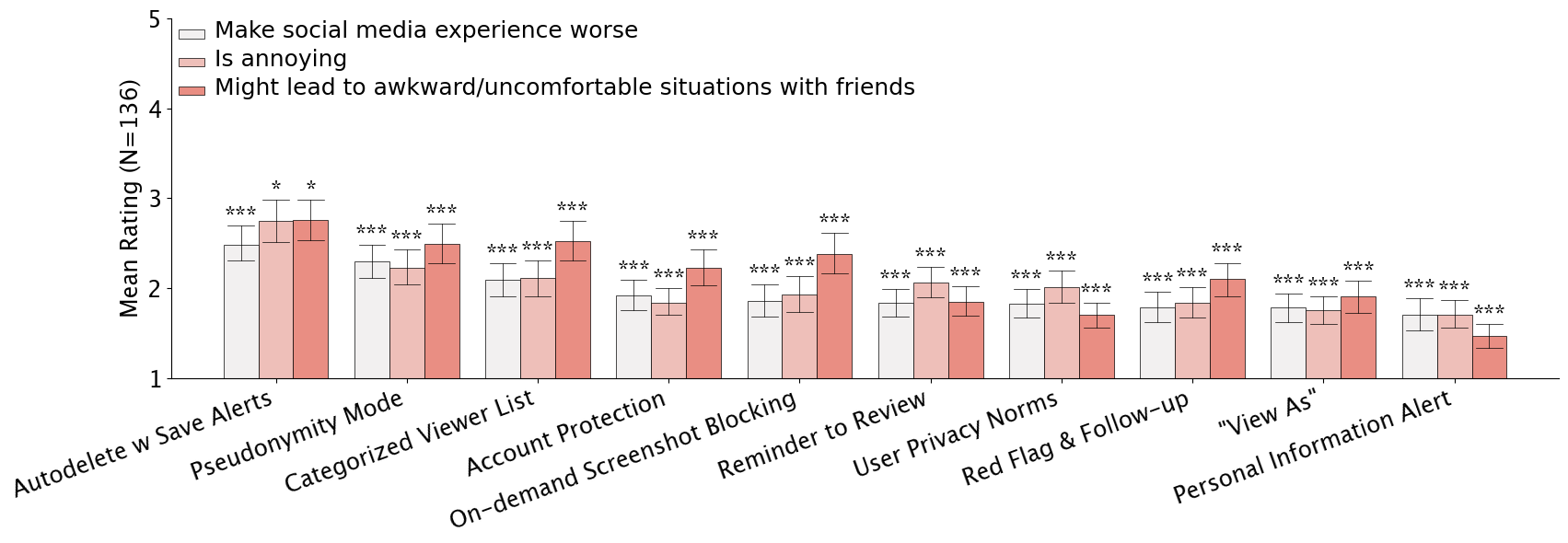}
    \caption{Level of potential trouble caused by features}
    \Description{Figure 2. (c) shows a bar graph of the mean rating of participants' concerns that features might make the social media experience worse (light red) or might lead to awkward/uncomfortable situations with friends (dark red). The ratings vary between 2 and 3, indicating moderate concern for both potential issues across all features.}
  \end{subfigure}

   \caption{Bar graphs illustrating the mean rating (N=136; 72 responses from private account owners and 64 from public account owners; a total of 118 participants responded) of the ten design prototypes in the design evaluation survey. Each bar illustrates the mean rating of participants' responses to the questions, with standard error bars. The significance levels of the one-sample t-tests against the hypothesis that the mean rating is a neutral score are denoted using asterisks over the bar, where one asterisk denotes \(p<.05\), two asterisks denote \(p<.01\), and three asterisks denote \(p<.001\).}
   \Description{Figure 4. describes the mean participant ratings (N=130) to the design evaluation survey questions for the 10 design approaches in bar plots. Standard error (SE) bars are also displayed on each bar. There are 3 sub-figures. Each bar in the charts has asterisks above it, indicating the level of statistical significance, with one asterisk denoting *p < .05, two asterisks for **p < .01, and three asterisks for ***p < .001. The presence of these asterisks suggests that the responses are statistically significant at various levels of confidence. All bars have at least one asterisk.}
   \label{tab:individual_evaluations}
\end{figure}

%% file: tables/overall_evaluations.tex
\begin{figure}
    \centering
    \includegraphics[width=0.9\linewidth]{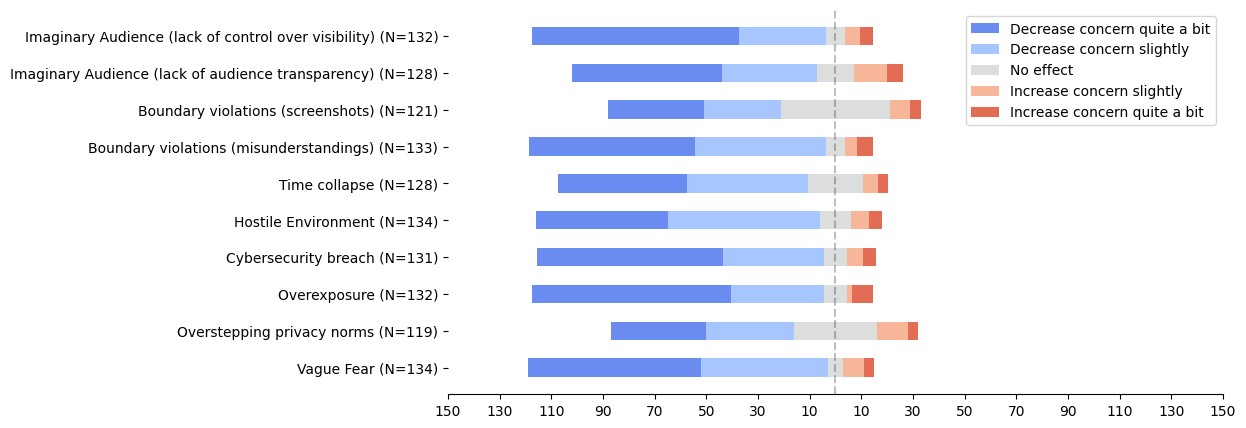}
    \caption{A diverging bar graph illustrates the ratings (N=136) for responses to the question, ``How would the features affect your [specific concern]?'' Although we initially identified seven categories of concern, we subdivided the fear related to ``imaginary audiences'' and ``boundary violations'' into two separate categories and added a category for vague fears, thereby totaling ten distinct types of concern.}
    \Description{The image is a diverging bar graph representing responses to how certain features affect specific concerns. The concerns are listed along the y-axis, each followed by the sample size (N) in parentheses. The concerns include: Imaginary Audience (lack of control over visibility) (N=132), Imaginary Audience (lack of audience transparency) (N=128), Boundary violations (screenshots) (N=121), Boundary violations (misunderstandings) (N=133), Time collapse (N=128), Hostile Environment (N=134), Cybersecurity breach (N=131), Overexposure (N=132), Overstepping privacy norms (N=119), Vague Fear (N=134); The x-axis is a horizontal scale ranging from -150 to 150, indicating the degree to which each concern is affected, with negative values representing a decrease in concern and positive values an increase. Each concern has two bars extending in opposite directions from the center, color-coded to represent the extent of the effect: Dark blue bars indicate a decrease in concern quite a bit; Light blue bars indicate a decrease in concern slightly; Grey bars; indicate no effect; Light red bars indicate an increase in concern slightly; Dark red bars indicate an increase in concern quite a bit. All bars extend predominantly to the left, indicating that the features were perceived to decrease the different types of fears.}
    \label{tab:overall_evaluations}
\end{figure}

%% file: sections/5_discussion.tex
\section{Discussion}
\subsection{Dysfunctional Fear and Privacy Norms}
Social media is central to teen communication, with privacy crucial for their safety and well-being. However, the focus on reducing privacy risks has often led to increased fear in teens' social media use. This fear is exacerbated by adult narratives that, while intended to promote vigilance, often resemble frightening folklore lacking specific details. Such stories contribute to a persistent perception of uncontrollable risks among teens, potentially leading to withdrawal from social media or dysfunctional anxiety that diminishes the quality of life without enhancing safety.

Current social media designs for mitigating teen privacy risks emphasize self-regulation, placing significant responsibility on individuals. However, teens must navigate multiple, often conflicting privacy norms across various contexts. Contextual Integrity theory~\cite{nissenbaum2004privacy} suggests that privacy is maintained when information flows appropriately according to contextual norms. Yet, for teens, adhering to these norms is often compromised by conflicting social pressures and platform designs that don't account for nuanced contexts.

Our study provides empirical evidence of dysfunctional fear in teens' privacy management. Teens expressed vague but persistent fears about audience control and feeling watched, coupled with concerns about potential negative social consequences of protective actions. They face dilemmas between privacy and social reach, experiencing significant stress while struggling to find effective solutions.

We found that social media platform design can significantly influence whether teens feel empowered to take constructive privacy-related actions. Our findings highlight a critical limitation in current approaches: the overemphasis on individual choice fails to account for the powerful influence of social contexts and norms that can override.

\input{tables/design_impact}

\subsection{A Design Agenda for Privacy Assurance}
Participants perceived tremendous potential and proposed numerous ideas for improving the status quo. Their primary agenda for change involved enhancing safety reassurance through system-level approaches, offering more options for users to tailor their privacy settings, and prioritizing privacy across the platform, thereby establishing privacy protection as the norm among users.

In our prototypes, based on interview data, we aimed to incorporate extensive user choice, allowing users to determine their own balance between utility and privacy. We enhanced existing features with nuanced controls, such as improved screenshot control and viewer lists supporting audience transparency. We also introduced new features like reminders to review past posts and reintroduced designs that teen participants were unfamiliar with, such as Facebook's ``View As'' function.

Based on our design evaluation results, we outline the following three key design recommendations:

\vspace{2mm}
\noindent\textbf{From Privacy Paralysis to Platform Responsibility.} Teens often experience dysfunctional fear of privacy issues on social media, leading to a perceived lack of self-efficacy. Paradoxically, this can result in them surrendering their privacy, believing issues to be inevitable. This suggests that the current model of individual user control is insufficient, highlighting the need for platforms to provide both protection and reassurance to effectively address teens' privacy concerns.

To enhance teenagers' confidence in managing privacy, social media platforms should promote privacy as a social norm by visibly prioritizing it in design and features. This includes explicitly communicating privacy expectations and providing varied protection options. Implementing features that offer tangible safety reassurance, such as follow-up notifications on reported incidents, can further bolster user confidence. Platforms should emphasize the importance of privacy during onboarding and throughout the user experience, empowering users to prioritize their privacy even when it conflicts with peer norms. For instance, explicitly stating that users should not feel obliged to accept all friend requests can help teens navigate complex social dynamics.

This approach shifts the narrative, making privacy protection a standard practice rather than an option for only the most cautious users. By providing a supportive environment, platforms can help teens navigate complex social interactions while maintaining privacy, acknowledging the significant influence of peer perceptions during adolescence. This strategy allows teens to learn safe social media navigation through managed risk-taking, as suggested by previous research, ultimately fostering a more privacy-conscious and empowered user base.

\vspace{2mm}
\noindent\textbf{Mapping Privacy-Utility Tensions: A Framework for Feature Evaluation.} Based on survey respondents' concerns, we identified four main areas of potential pitfalls of the privacy features: negative impact on other users, restriction of social media utility, user burden (e.g., features triggered without active engagement), and community-dependency (where individual engagement alone is insufficient to reduce privacy risks). We analyzed each prototype against these criteria, summarizing the results in Table \ref{tab:impact}. As anticipated, features without these drawbacks---Personal Information Alert and Red Flag \& Follow-up---were most positively received. Conversely, Auto-delete (with save), which exhibited all four issues, was the least favored.

These findings underscore that effective designs must balance privacy enhancement with utility while considering impacts on user experience, social dynamics, and perceived effectiveness. First, privacy is secondary to engagement for teens on social media. Features that significantly impact utility or impose a high user burden are less likely to be embraced. Second, teens are highly attuned to peer dynamics and social norms, preferring features that do not burden their friends. Third, community-dependent features may be perceived as less effective, potentially due to their limited impact on individual self-efficacy or agency. These highlight the delicate balance required in designing privacy features for teens: they must enhance privacy without compromising the core social and functional benefits that draw teens to these platforms.

\vspace{2mm}
\noindent\textbf{Beyond Binary Privacy: Public Accounts' Privacy Protection Needs.} Our study revealed that teens have varied privacy requirements, particularly evident in the contrast between private and public account users. While private account holders prioritize strict privacy controls, public account users, such as aspiring artists, seek a balance between visibility and protection. However, public account owners also reported twice as much experience with a dysfunctional fear of privacy risks than private account owners. 

To address this spectrum of needs, participants proposed design improvements that offer nuanced privacy options. For public accounts, suggestions included disconnecting personal information from public visibility, systematically filtering out users who fail to meet community standards, and enhancing interaction transparency through better-organized viewer lists. These features aim to enable public account holders to engage with a larger audience without fully compromising their privacy, demonstrating the need for flexible privacy solutions that cater to diverse user goals.

Participants also advocated for a range of privacy-enhancing features applicable to all users, reflecting the individual variation in privacy concerns. These included alerts for potential privacy compromises, systematic visibility controls like auto-delete or screenshot blocking, and the ability to view one's profile from others' perspectives. Recognizing that the perceived necessity and potential drawbacks of these features vary among individuals, participants emphasized the importance of user choice. They preferred the option to toggle features on and off, with privacy-focused settings as the default. This approach not only accommodates individual preferences but also reinforces privacy as the norm while allowing users to customize their experience.

\subsection{Limitations and Future Work}
We did not conduct a thematic analysis of the survey's free-response questions, so it is unclear which reactions were most prevalent among respondents. Additionally, our survey did not measure the varying intensities of different fears (e.g., whether some types of fear are more likely to lead to dysfunctional behavior than others), nor did it explore how different design ideas might affect each type of fear. We also did not investigate the distinct responses of public and private account holders to various fears and design ideas in our study. Further, our co-design interview participants were not limited to teens with a dysfunctional fear of privacy issues, nor were our design evaluation survey respondents. While this means that we did not explore features specifically targeting dysfunctional fear, the findings from our study are transferable to designs for dysfunctional fear. This is because the teen participants indicated that our features reduce fear and risks, critical elements in decreasing dysfunctional fear.

While we received numerous responses outlining the potential drawbacks of each feature, we utilized prototypes, meaning that in real-world and long-term scenarios, these features might encounter unforeseen issues. Using specific features as examples for each design idea made it easier for teens to understand and provide detailed feedback. However, there are multiple ways to implement these design ideas, and our findings may not be generalizable to all implementations. Although our design directions are intended to be broader concepts that can be adapted to different implementations, this is a limitation of our study. Another assumption in our prototypes was the feasibility and accuracy of user reporting systems, which is not currently a given. This assumption was necessary as solving this problem was beyond the scope of our research.

A significant challenge, consistent with prior research, is addressing fear in the context of real-life peer dynamics on social media disclosure. Future research should delve into how the content shared by teens might impact their social interactions (e.g., fear of judgment by peers based on a single post) and explore more effective solutions for this issue.

%% file: tables/design_impact.tex
\begin{table*}[!t]
\caption{Assessment of potential drawbacks and/or trade-offs for each prototype.}
\label{tab:impact}
    \begin{tabularx}{\textwidth}{@{} >{\raggedright\arraybackslash}p{4.2cm} >{\raggedright\arraybackslash}p{2cm} >{\raggedright\arraybackslash}p{2cm} >{\raggedright\arraybackslash}p{2cm} >{\raggedright\arraybackslash}X@{}}
    \toprule
    \textbf{Feature Name} & \textbf{Impact on Other Users} & \textbf{Utility Restriction} & \textbf{User Burden} & \textbf{Community-Dependent} \\
    \midrule
    Auto-delete (with save) & Yes & Yes & Yes & Yes \\
    Reminder to review & No & No & Yes & No \\
    Account protection & No & Yes & No & No \\
    On-demand screenshot blocking & Yes & Yes & No & Yes \\
    Personal information alert & No & No & No & No \\
    User privacy norms & No & No & No & Yes \\
    Categorized viewer list & No & No & Yes & No \\
    ``View as'' & No & No & Yes & No \\
    Red flag \& follow-up & No & No & No & No \\
    Pseudonymity mode & No & Yes & No & Yes \\
    \bottomrule
    \end{tabularx}
\end{table*}

%% file: sections/6_conclusion.tex
\section{Conclusion}
We conducted co-design interviews with 19 teens and a design evaluation survey with 136 U.S. teen participants. The study revealed that teens experience significant dysfunctional privacy fears on social media, with 28.1\% of public account users and 15.3\% of private account users reporting fears that diminish their quality of life without enabling constructive protective actions. These fears stem from a perceived lack of control over audience reach, hostile online environments, and uncertainty about privacy norms.

Our work suggests that platforms should take greater responsibility for establishing privacy as a social norm rather than focusing solely on individual vigilance and restrictive measures. The ten design prototypes we evaluated demonstrate concrete ways to empower users while reducing unnecessary anxiety. The features that were perceived as most effective shared common characteristics: They minimized impact on other users, had low trade-offs with existing functionality, required minimal user effort, and functioned independently of community behavior. Particularly successful examples included alerts about potential privacy exposures and systematic follow-up on user reports.

The implications of this work extend beyond specific features to suggest a broader shift in how platforms approach teen privacy. Rather than emphasizing individual responsibility and vigilance, social media platforms should work to establish privacy protection as a community norm through system-level design choices. This involves not only providing robust privacy controls but also creating an environment where privacy-conscious behavior is explicitly valued and supported. Future research should continue to explore how platform design can help transform vague privacy anxieties into constructive privacy practices, enabling teens to fully participate in online social spaces while maintaining appropriate boundaries.

%% file: sections/7_appendix.tex
\clearpage
\appendix
\label{appendix}

\section{Demographics Data}
\label{appendix-A}
\input{tables/demographics}
\vspace{3mm}

\section{Identifying Dysfunctional Worry}
\label{appendix-B}
\input{tables/worry_groups}
\vspace{3mm}

\section{UMAP~\cite{mcinnes2018umap} Clusters of Open-ended Responses in Follow-up Survey}
\label{appendix-C}
\input{tables/clusters}

%% file: tables/demographics.tex
\begin{table}[!h]
\centering
\caption{Demographics of the Co-design Survey Participants (N=19)}
\Description{Table 2. provides demographic information about the participants involved in the co-design study. It includes gender distribution, average age, and racial background of the participants. It describes the platforms that participants share content on frequently and comfortably. There are 2 columns with the categories in the first column on a grey background with black text, and the data in the second column on a white background with black text. }
\label{tab:phase2_demographics}
\begin{tabular}{|p{4cm}|p{9cm}|}
\hline
\cellcolor[HTML]{C0C0C0} Gender identity & Girls (63\%), Boys (26\%), Non-binary or third gender (5\%), Boy and non-binary or third gender (5\%) \\
\hline
\cellcolor[HTML]{C0C0C0} Age & 13 (10.5\%), 14 (10.5\%), 15 (15.8\%), 16 (21.1\%), 17 (21.1\%), 18 (21.1\%) \\
\hline
\cellcolor[HTML]{C0C0C0} Race & White (53\%), Asian or Asian-American (32\%), Black or African American (11\%), White and Asian or Asian-American (5\%) \\
\hline
\cellcolor[HTML]{C0C0C0} Hispanic or Latin-American Origin & No (84\%), Yes (16\%) \\
\hline
\cellcolor[HTML]{C0C0C0} Social Media Usage & Instagram (100\%), BeReal (84\%), Snapchat (74\%), Twitter (68\%), TikTok (63\%), Discord (11\%), Reddit (11\%), Tumblr (5\%) \\
\hline
\cellcolor[HTML]{C0C0C0} Platforms Where Participants Reported Sharing Most Frequently & Instagram (42\%), BeReal (32\%), Snapchat (11\%)), Twitter (5\%), YouTube (5\%), Discord (5\%) \\
\hline
\cellcolor[HTML]{C0C0C0} Platforms Where Participants Reported Sharing Most Comfortably & Instagram (36\%), BeReal (26\%), Snapchat (16\%), Discord (16\%), Pinterest (5\%) \\
\hline
\end{tabular}
\end{table}

\begin{table}[!h]
\centering
\caption{Demographics of the Follow-up (Design Evaluation) Survey Participants (N=136)}
\Description{Table 3. contains information for the participants in the Follow-up (Design Evaluation) Survey. The table provides insights into the gender identity, mean age, race, Hispanic or Latin-American origin, and social media usage of the participants. These demographic details are essential for understanding the composition of the survey participants and are crucial for interpreting the survey results. There are 2 columns with the categories in the first column on a grey background with black text, and the data in the second column on a white background with black text. }
\label{tab:phase3_demographics}
\begin{tabular}{|p{4cm}|p{9cm}|}
\hline
\cellcolor[HTML]{C0C0C0} Gender identity & Girls (59.6\%), Boys (33.8\%), Non-binary or third gender (2.9\%), Girl and non-binary or third gender (1.5\%), Boy, girl, and non-binary or third gender (1.5\%), Prefer Not to Disclose (0.7\%) \\
\hline
\cellcolor[HTML]{C0C0C0} Age & 13 (0\%), 14 (5.1\%), 15 (18.4\%), 16 (18.4\%), 17 (29.4\%), 18 (28.7\%) \\
\hline
\cellcolor[HTML]{C0C0C0} Race & White (47.8\%), Black or African American (25.7\%), Asian or Asian-American (20.6\%), White and Black or African American (1.5\%), White and Asian or Asian-American (0.7\%), White and American Indian or Alaska Native (1.5\%), Black or African-American, American Indian or Alaska Native (1.5\%), Asian or Asian-American, Black or African-American (0.7\%), Other (0.7\%) \\
\hline
\cellcolor[HTML]{C0C0C0} Hispanic or Latin-American Origin & No (83.1\%), Yes (16.9\%) \\
\hline
\cellcolor[HTML]{C0C0C0} Social Media Usage & Instagram (Real) (85.3\%), Snapchat (55.9\%), TikTok (65.4\%), BeReal (38.2\%), Instagram (Spam) (39.7\%),  Twitter (40.4\%)

Other (7.9\%): Reddit (2.9\%), Youtube (1.5\%), Pinterest (0.7\%), Locket (0.7\%), Lemon8 (0.7\%), Whatsapp (0.7\%), Zeeme (0.7\%) \\
\hline
\end{tabular}
\end{table}

%% file: tables/worry_groups.tex
The following is the algorithm for identifying ``Unworried'' vs. ``Functional Worry'' vs. ``Dysfunctional Worry'' groups, adopted from Jackson and Gray (2010)~\cite{jackson2010functional}. We first ask the following five questions as part of the worry group survey:

\begin{enumerate}[label=\Roman*.]
    \item Overall, how much do you worry about privacy-related issues on the [public/private] account on [platform name]? (1: Not at all worried; 2: Not very worried; 3: Fairly worried; 4: Very worried)
    \item What steps do you take to protect yourself on your [public/private] account on [platform name]? (0: Never; 1: Occasionally; 2: Often; 3: Always, on five different ways of privacy protection measures)
    \item To what extent do you feel these measures are effective in protecting your privacy on your [public/private] account on [platform name]? (0: Not at all; 1: A little; 2: Moderately; 3: Quite a bit; 4: Very much)
    \item How much do your concerns about privacy issues on your [public/private] account on [platform name] affect your overall quality of life? (0: Not at all; 1: A little; 2: Moderately; 3: Quite a bit; 4: Very much)
    \item How much do the steps you take to protect your privacy on [public/private] account on [platform name] affect your quality of life? (0: Not at all; 1: A little; 2: Moderately; 3: Quite a bit; 4: Very much)
\end{enumerate}

\begin{table*}[!h]
\caption{Coding response to questions in the worry group survey.}
\label{tab:worry-coding}
    \begin{tabular}{p{0.08\linewidth} p{0.28\linewidth} p{0.53\linewidth}}
    \toprule
    \textbf{Code Label} & \textbf{Code Description} & \textbf{Coded responses to survey questions} \\
    \midrule
        (1a) & Not worried about crime & ``1: Not at all worried'' or ``2: Not very worried'' to Question I. \\
        (1b) & Worried about crime & ``3: Fairly worried'' or ``Very worried'' on Question I. \\
        \hline
        (2a) & Worry has no impact on quality of life & ``0: Not at all'' or ``1: A little'' on Question IV. \\
        (2b) & Worry has impact on quality of life & ``2: Moderately'' or ``3: Quite a bit'' or ``4: Very much'' on Question IV. \\
        \hline
        (3a) & Don't take precautions & Average of five sub-scores to Question II. rounds to 0. \\
        (3b) & Take precautions & Average of five sub-scores to Question II. rounds to 1, 2, or 3. \\
        \hline
        (4a) & Precautions have no impact on safety & ``0: Not at all'' or ``1: A little'' on Question III. \\
        (4b) & Precautions have impact on safety & ``2: Moderately'' or ``3: Quite a bit'' or ``4: Very much'' on Question III. \\ 
        \hline
        (5a) & Precautions have no impact on quality of life & ``0: Not at all'' or ``1: A little'' on Question V. \\
        (5b) & Precautions have impact on quality of life & ``2: Moderately'' or ``3: Quite a bit'' or ``4: Very much'' on Question V.\\
        \hline
    \bottomrule
    \end{tabular}
\end{table*}

Based on the coded responses above, we categorize respondents into three groups---``Unworried,'' ``Functional Worry,'' and ``Dysfunctional Worry'' based on the conditions below:
\begin{itemize}
    \item Unworried Group: (1a)
    \item Functional Worry Group: \{(1b) \&\& (2b) \&\& (3b) \&\& (4b) \&\& (5a)\} || \{(1b) \&\& (2a)\}
    \item Dysfunctional Worry Group: \{(1b) \&\& (2b) \&\& (3a)\} || \{(1b) \&\& (2b) \&\& (3b) \&\& (4a)\} || \{(1b) \&\& (2b) \&\& (3b) \&\& (4b) \&\& (5b)\}
\end{itemize}

%% file: tables/clusters.tex
\newcolumntype{L}[1]{>{\raggedright\let\newline\\\arraybackslash\hspace{0pt}}p{#1}}

\begin{center}
\setstretch{1.05}
\begin{longtable}{L{1.8cm}L{2.2cm}L{8.9cm}}
\caption{UMAP cluster labels and example quotes for each of the 10 prototypes used in the follow-up design evaluation survey. The numbers in parenthesis in the ``Cluster Label'' column indicate the number of responses that were classified into the cluster.} \label{tab:clusters} \\

\hline \multicolumn{1}{c}{\textbf{Feature}} & \multicolumn{1}{c}{\textbf{Cluster Label}} & \multicolumn{1}{c}{\textbf{Example Quotes}} \\ \hline 
\endfirsthead

\multicolumn{3}{c}%
{{\bfseries \tablename\ \thetable{} -- continued from previous page}} \\
\hline \multicolumn{1}{c}{\textbf{Feature}} & \multicolumn{1}{c}{\textbf{Cluster Label}} & \multicolumn{1}{c}{\textbf{Example Quotes}} \\ \hline 
\endhead

\hline \multicolumn{3}{r}{{Continued on next page}} \\ \hline
\endfoot

\hline \hline
\endlastfoot

    Account Protection (Likes) & 0: User Restriction (33) & ``\textit{I like that you can restrict certain users that have a certain amount of red flags}'' (R6); ``\textit{I like that this feature could prevent suspicious users or bots from interacting with your content.}'' (R51) \\
    & 1: Extra Security (34) & ``\textit{It lets you have extra security measures}'' (R104); ``\textit{i like the security idea. it gives me a sense of comfort as i believe the extra security layers is an advantage}'' (R123) \\
    & 2: Customizability (66) & ``\textit{It gives the user more options on how to customize their accounts privacy}'' (R108); ``\textit{I love the specificity and variety in the options for who can view your profile.}'' (R119) \\
    \hline
    Account Protection (Dislikes) & 1: Loopholes (32) & ``\textit{How do you filter for people who are very new and suspiciously active? What if someone creates a new account to follow you?}'' (R32); ``\textit{I think it's good, but people can still create new accounts to stalk.}'' (R58)\\
    & 2: Complexity (16) & ``\textit{It looks complicated}'' (R87); ``\textit{Too many sub features}'' (R90) \\
    & 3: Uncertain Criteria (28) & ``\textit{i dont know exactly how meaningful the red flags are- how are they given?}'' (R79); ``\textit{Im just concerned about how red flags will be handed out and how accurate it will be. Also countermeasures and ways to remove red flags that were placed unjustifiably.}'' (R116) \\
    & 0, 4, 5: None (45) & ``\textit{No complaints! :)}'' (R21); ``\textit{I do not dislike anything.}'' (R47) \\
    \hline
    Categorized Viewer List (Likes) & 0, 2: Reviewing Followers (114) & ``\textit{The fact I can see who I've been talking to and who I haven't been talking to also it might help filter out fake accounts}'' (R93); ``\textit{It sorts your friends based on how much you chat with them. So, you can keep your real buddies at the top and decide who stays on your follower list.}'' (R107) \\
    & 1: Contentment (18) & ``\textit{I love this!!}'' (R35); ``\textit{It's interesting}'' (R102) \\
    \hline
    Categorized Viewer List (Dislikes) & 2: Too much Transparency (54) & ``\textit{I think it sacrifices the privacy of the viewers a little bit: some people might not want others to know how often they view/interact with their content.}'' (R1); ``\textit{I dont like the idea of having a list of people who view your stuff, it can become an obsession issue where people are constantly checking who viewed their posts and may use it to cause drama or rumors. It's unnecessary and causes drama and mental health issues.}'' (R36) \\
    & 3: Excessive (15) & ``\textit{seems a little extra}'' (R37); ``\textit{Too complex}'' (R89) \\
    & 0, 1: None (51) & ``\textit{Great feature. I dislike nothing.}'' (R27); ``\textit{Nothing really}'' (R109) \\
    \hline
    Auto-delete with Save (Likes) & 0, 5: Record Control (58) & ``\textit{i like that i can control how long my chats are present, minimizing my digital footprint}'' (R19); ``\textit{I like that there isnt a record of everything you have said in the past.}'' (R62) \\
    & 2: Mutual Transparency (35) & ``\textit{I like the second feature where it notifies both sides if a screenshot is taken.}'' (R124); ``\textit{It tells me in advance that saving will notify the others. And can save as an image.}'' (R23) \\
    & 1, 3, 4: None (37) & ``\textit{nothing}'' (R30); ``\textit{I don't like this feature.}'' (R111) \\
    \hline
    Auto-delete with Save (Dislikes) & 1: Time Period (25) & ``\textit{3 days is long enough to show it to other people IRL and have you conversation spread without you knowing.}'' (R77); ``\textit{Yes it seems like 3 days is too early for the message to be deleted}'' (R78) \\
    & 2: Lost Record (65) & ``\textit{Being able to permanently erase chats can clear evidence of crimes committed on the app and so on.}'' (R42); ``\textit{I forget conversations from a couple days ago easily, so this increases the likelyhood of me forgetting and never remembering.}'' (R109) \\
    & 0: None (33) & ``\textit{I honestly have no complaints about this feature!}'' (R20); ``\textit{I don't dislike anything about this feature.}'' (R43) \\    
    \hline
    On-demand Screenshot Blocking (Likes) & 0, 1, 4: Control over Sharing (108) & ``\textit{Good way to prevent unwanted sharing of a post meant specifically for a certain audience. Nice!}'' (R24); ``\textit{I really like this feature because it makes sure the content I share will be less likely to be shared with others; it makes it harder for the content to go places.}'' (R56) \\
    & 2, 3: Contentment (26) & ``\textit{I really like this!}'' (R35); ``\textit{Looks amazing}'' (R99) \\
    \hline
    On-demand Screenshot Blocking (Dislikes) & 0: No Notification (25) & ``\textit{I would make it to where (like snapchat) it lets the owner of the account know who, and when they screenshotted, i think that could help prevent a lot of mental health issues and bullying.}'' (R33); ``\textit{Maybe a sort of notification can be sent to me after someone tries to take a screenshot, probably with the person’s name asking me if I would like to let them take the shot}'' (R87) \\
    & 3: Loopholes (25) & ``\textit{It prevents some unconsensual sharing, however it isn't stopping people from showing other in person or finding another method.}'' (R17); ``\textit{Unfortunately, people will find their way around it, like using a different phone to take a picture of it.}'' (R105) \\
    & 1, 2: None (69) & ``\textit{I love it as it is :)}'' (R37); ``\textit{I do not dislike anything.}'' (R113) \\
    \hline
    Reminder to Review (Likes) & 0: Contentment (11) & ``\textit{This is amazing}'' (R73); ``\textit{It is cool.}'' (R109) \\
    & 1: Self-Reflection (115) & ``\textit{WOW! This is a phenomenal feature. I tend to go back to my social media and reconsider or delete posts that reflect "the old me" or are just too old for my liking.}'' (R23); ``\textit{It reminds you to check your old posts, so you make sure they still fit who you are now. Keeps your online self in sync with the real you.}'' (R107) \\
    & 2: None (6) & ``\textit{nothing really}'' (R52); ``\textit{Nothing}'' (R108) \\
    \hline
    Reminder to Review (Dislikes) & 0: Unnecessary (6) & ``\textit{I think it's unnecessary}'' (R38); ``\textit{I just don't think it's necessary for me.}'' (R74) \\
    & 3: Annoying (28) & ``\textit{it seems annoying and kind of pointless}'' (R3); ``\textit{I don't like the part where it gives a reminder, and I would hope that this feature doesn't show every single time you post because it would be annoying.}'' (R18) \\
    & 5: Inflexibility (40) & ``\textit{I don't like that it only gives the option to delete after 24 hours and not a chosen time period, although that might be just for the example.}'' (R17); ``\textit{I dislike how the opposing option to posting is to delete permanently. I feel a better option to deleting would be an option to private/archive the post so it's not available to the public but still to the user.}'' (R25) \\
    & 1, 2, 4: None (44) & ``\textit{I dont dislike anything about it.}'' (R5); ``\textit{none}'' (R69) \\ 
    \hline
    Red Flag and Follow-up (Likes) & 0, 2: Reassurance (37) & ``\textit{I like that you are able to receive updates about reports, along with the details of the punishment.}'' (R102); ``\textit{I love how it allows you to know if your concern has been heard instead of wondering if it has or hasn't been taken action for.}'' (R128) \\
    & 1: Contentment (20) & ``\textit{Everything about the features}'' (R7); ``\textit{Love it}'' (R36) \\
    & 3, 4: Safer Environment (77) & ``\textit{I like this because it allows users of the app to hold each other accountable and create a safer environment.}'' (R57); ``\textit{I like that it is included because it should be standard with any sharing platform. It allows users to take care of themselves by reporting people/content that make them feel unsafe.}'' (R115) \\
    \hline
    Red Flag and Follow-up (Dislikes) & 0: Lack of Variation (26) & ``\textit{As it shows here it says banned for a month from the platform. I'd hope that there would be varying degrees of punishments.}'' (R24); ``\textit{I think it seems pretty good. I would make sure that the ban period/consequence for the flag varies depending on how bad the flag was.}'' (R68) \\
    & 2: Misuse (34) & ``\textit{This feature can be abused and people can start reporting things just because they don't like the person that posted it, etc. Also, a lot of times the app ignores reports that are serious.}'' (R101); ``\textit{The only thing I have to say about this feature is that I would hope that others wouldn't misuse the report button just to get others banned unreasonably.}'' (R114) \\
    & 1, 3: None (56) & ``\textit{I dont dislike it! Its quite nice}'' (R4); ``\textit{Nothing I dislike about this feature}'' (R71) \\
    \hline
    Pseudonymity Mode (Likes) & 0: Audience Control (16) & ``\textit{I like how I can choose who can see my profile and who can't}'' (R27); ``\textit{I like the control over who can interact/view my account}'' (R50) \\
    & 1: Contentment (22) & ``\textit{It's definitely helpful}'' (R9); ``\textit{It's good feature}'' (R101) \\
    & 2, 5, 6: Advanced Anonymity (70) & ``\textit{I like how you can customize what you appear as to other people; it adds a layer of anonymity and acts as a preventative measure against potential harassers.}'' (R59); ``\textit{How it truly makes it you private.}'' (R63) \\
    & 3, 4: None (23) & ``\textit{I don't like it.}'' (R55); ``\textit{I don't think I really like this}'' (R116) \\
    \hline
    Pseudonymity Mode (Dislikes) & 1: Misuse and Challenges in Connecting (77) & ``\textit{I don't like this feature because it could be used in the wrong way. For example, someone could use this feature to bully someone by hiding their profile from the victim.}'' (R47); ``\textit{It can be hard to figure out who someone is. It also can keep people from finding potential friends online.}'' (R20) \\
    & 0, 2: None (48) & ``\textit{I don't have any specific issues with it.}'' (R27); ``\textit{There's nothing I dislike}'' (R61) \\
    \hline
    Personal Information Alert (Likes) & 0, 1, 2: Error Prevention and Awareness (134) & ``\textit{Ooh I like this, because sometimes people can miss things, so if the app's able to catch things like that before the post is even sent out, that'd be super helpful.}'' (R130); ``\textit{I like that it can help people realize that their post may contain sensitive information that really shouldn't be available to see. Especially for people who post more impulsively.}'' (R68) \\
    \hline
    Personal Information Alert (Dislikes) & 0: Inaccuracy and Inflexibility (53) & ``\textit{I do wonder if the feature would always work properly every time(picking up the information or failing to pick up the information).}'' (R57); ``\textit{Maybe they should have a studio feature where you can blur/edit things in the app so you don't have to redo the whole post, instead you can go into the studio for one specific photo in the post (if multiple)} (R31) \\
    & 1, 2: None (60) & ``\textit{Nothing to dislike}'' (R66); ``\textit{I can't think of anything.}'' (R99) \\
    \hline
    User Privacy Norms (Likes) & 0, 4: Explicitness (51) & ``\textit{I like that it spells out the guidelines for you and makes it easy to understand how to be safe on the internet.}'' (R19); ``\textit{I like that it has pretty concise bullet points that are easily understandable for everyone.}'' (R67) \\ 
    & 1: Contentment (7) & ``\textit{The features is very useful}'' (R77); ``\textit{It’s a great feature}'' (R97) \\
    & 2, 3: Establishing Norm (50) & ``\textit{I love that it makes being respectful normal instead of making out people who want privacy and respect online to be weird.}'' (R45); ``\textit{I like that it's straight out there: being privacy-conscious isn't odd, but rather smart.}'' (R129) \\
    & 5: Safety Reminders (25) & ``\textit{It is a reminder that the internet can be a dangerous place, and privacy should be prioritized}'' (R10); ``\textit{I like that it gives people a reminder of how to keep themselves safe online.}'' (R64) \\
    \hline
    User Privacy Norms (Dislikes) & 2: Vagueness and Potential Ignorance (23) & ``\textit{I think some of the guidelines are conceptual/not really concrete so some users may not really read thoroughly through them.}'' (R50); ``\textit{Most people do not really take the time to read such guidelines or terms when they use an app, so this might go ignored by many people.}'' (R108) \\
    & 5: Annoying (17) & ``\textit{seems a bit annoying, i wouldnt even read all those words to be honest id just click skip}'' (R30); ``\textit{It depends on how often the reminder pops up but if it's often it might become annoying.}'' (R11) \\
    & 6: Lengthy (23) & ``\textit{It’s very wordy and I have a feeling lots of teenagers would just skip right past it and not read it}'' (R32); ``\textit{some might find it too long to read, maybe summarizing a bit would help}'' (R91) \\
    & 0, 1, 3, 4: None (55) & ``\textit{I have no issues with the feature}'' (R33); ``\textit{Nothing at all}'' (R102) \\
    \hline
    “View As” (Likes) & 0, 3: Reassurance (88) & ``\textit{This would be extremely useful to make sure one does not have anything exposed that they would not want to be.}'' (R20); ``\textit{You can check how your profile looks to others, making sure you share only what you want. It's like a double-check for your online image.}'' (R109) \\
    & 1: Contentment (17) & ``\textit{I like it for the most part.}'' (R66); ``\textit{It's interesting}'' (R75) \\
    & 2: Different POV (29) & ``\textit{It allows me to see how friends, peers, and maybe strangers would see my account from their point of view/perspective.}'' (R44); ``\textit{It gives you better insight on how others see your account, without you having to assume or ask another person to check for you.}'' (R126) \\
    \hline
    “View As” (Dislikes) & 0, 3: Excessive and Unnecessary (32) & ``\textit{I think this feature can easily become quite complicated, and especially if multiple users in the following list are being shown different parts.}'' (R25); ``\textit{It's kind of unneeded because what others see on your account is pretty much what you can see on your end}'' (R51) \\
    & 1, 2, 4, 5: None (79) & ``\textit{don’t dislike anything}'' (R5); ``\textit{I have nothing negative to say.}'' (R13) \\
\end{longtable}
\end{center}